\documentclass[fleqn,usenatbib]{mnras}

\usepackage{newtxtext,newtxmath}

\usepackage[T1]{fontenc}

\DeclareRobustCommand{\VAN}[3]{#2}
\let\VANthebibliography\thebibliography
\def\thebibliography{\DeclareRobustCommand{\VAN}[3]{##3}\VANthebibliography}

\usepackage{graphicx}	
\usepackage{amsmath}	
\usepackage{makecell}
\usepackage{graphicx}
 \usepackage{caption}
\usepackage{subcaption}
\usepackage{nicefrac,xfrac}
\usepackage{array, multirow}

\title[Dynamical Density Profiles]{From particles to orbits: precise dark matter density profiles\\ using dynamical information}

\author[C. Muni et al.]{
Claudia Muni$^{1}$\thanks{E-mail: claudia.muni.21@ucl.ac.uk},
Andrew Pontzen$^{1}$,
Jason L. Sanders$^{1}$,
Martin P. Rey$^{2}$,
Justin I. Read$^{3}$,
Oscar Agertz$^{4}$
\\
$^{1}$Department of Physics $\&$ Astronomy, University College London, Gower Street, London WC1E 6BT, UK\\
$^{2}$Sub-department of Astrophysics, University of Oxford, DWB, Keble Road, Oxford OX1 3RH, UK\\
$^{3}$School of Mathematics $\&$ Physics, University of Surrey, Guildford, GU2 7XH, UK\\
$^{4}$Lund Observatory, Division of Astrophysics, Department of Physics, Lund University, Box 43, SE-221 00 Lund, Sweden\\
}

\date{Accepted XXX. Received YYY; in original form ZZZ}

\pubyear{2023}

\begin{document}
\label{firstpage}
\pagerange{\pageref{firstpage}--\pageref{lastpage}}
\maketitle

\begin{abstract}
We introduce a new method to calculate dark matter halo density profiles from simulations. Each particle is ‘smeared’ over its orbit to obtain a dynamical profile that is averaged over a dynamical time, in contrast to the traditional approach of binning particles based on their instantaneous positions.
The dynamical and binned profiles are in good agreement, with the dynamical approach showing a significant reduction in Poisson noise in the innermost regions. 
We find that the inner cusps of the new dynamical profiles continue inward all the way to the softening radius, reproducing the central density profile of higher resolution simulations within the 95$\%$ confidence intervals, for haloes in virial equilibrium. 
Folding in dynamical information thus provides a new approach to improve the precision of dark matter density profiles at small radii, for minimal computational cost.
Our technique makes two key assumptions: that the halo is in equilibrium (phase mixed), and that the potential is spherically symmetric. We discuss why the method is successful despite strong violations of spherical symmetry in the centres of haloes, and explore how substructures disturb equilibrium at large radii.
\end{abstract}




\section{Introduction} \label{introduction}

The observationally inferred density distribution of dark matter in haloes around galaxies offers a crucial hint as to the nature of the elusive substance. However, the observations must be carefully compared with theoretical predictions based largely on numerical simulations (for reviews see e.g. \citealt[][]{Frenk_2012, Vogelsberger_2020, Angulo_2022}). 
Dark-matter-only (DMO) simulations have shown that the spherically-averaged density profiles of haloes in the Cold Dark Matter (CDM) paradigm follow approximately the Navarro-Frenk-White (NFW) profile \citep[][]{Dubinski_1991, Navarro_1996, Navarro_1997, Dutton_2014} described by a divergent cusp ($\rho \sim r^{-1}$) at small radii, and by a steeper power law ($\rho \sim r^{-3}$) in the outer regions. 
The NFW profile has two free parameters which may be fitted to the density structure of simulated haloes for most of the radial extent, but the fit becomes poor in the  innermost parts and in the outskirts of the haloes \citep[e.g.][]{Navarro_2004, Diemer_2014, Fielder_2020, Wang_2020_N, LucieSmith22}.

Over time, a variety of fitting functions have been proposed to better represent the profile's inner slope, such as Einasto models \citep[][]{Einasto_1965, Chemin_2011} or other forms of double-power law \citep[e.g.][]{Hernquist_1990, Burkert_1995, Zhao_1996, Salucci_2007, Hague_2013, Oldham_2016, Hayashi_2020}.
However, the central regions of the profiles remain notoriously difficult to probe due the finite number of particles and consequent need to `soften' the potential \citep[e.g.][]{Power_2003, Diemand_2004, Dehnen_2011}, causing the cusp to be numerically flattened \citep[e.g.][]{Navarro_1996, Ghigna_2000, Fukushige_2001, Wang_2020_N}. 
Constraining the central asymptotic behaviour of the profile therefore remains largely dependent on the number of particles concentrated at small radii.

While the focus in the present work will be on DMO simulations, we note that when baryons are added into simulations, effects such as supernova feedback and enhanced dynamical friction can cause the central cusp to turn into a flattened density `core' \citep[e.g.][]{Navarro_1996a, Read_2005, Pontzen_2012, Read_2016, El-Zant_2001, Nipoti_2014, Del_Popolo_2016, Orkney_2022}. Ultimately, understanding the predicted distribution of dark matter does require such baryonic simulations, especially since there are strong indications of flattened central cores in observations; see e.g. \citet{Flores_1994, deBlok_2001, Marchesini_2002, Battaglia_2008, Walker_2011, Oh_2015, Read_2017, Read_2019, Zoutendijk_2021, DeLeo_2023}, or for countering views see  \citet{Pineda_2016, Genina_2017, Oman_2018}.  The focus in the present work is nonetheless on understanding how DMO predictions can be improved and better understood; we will consider baryonic effects in a future paper.

In the outskirts of haloes, density profiles scatter significantly due to the presence of surrounding substructures and the out-of-equilibrium dynamics of accreting material. For instance, the caustics generated by the infalling particles on their first apocentre passage sets the scale for the splashback radius, which creates an observable signature in the outer regions of halo profiles \citep[][]{Diemer_2014, Adhikari_2014, More_2015, Shin19Splashback}. Recently, \cite{LucieSmith22} showed that a good fit to the diversity of halo profiles out to two virial radii can be obtained using only three free parameters (i.e., one additional parameter is sufficient to capture the diversity of these outer regions). This relatively simple behaviour may be linked to the typical orbits on which material accretes into a halo, further motivating a study of how the instantaneous profile relates to a dynamically-generated equilibrium profile  \citep[e.g.][]{Diemer_2022, Diemer_2022b, Shin_2023}.

In this work, we present and study a method to calculate dark matter density profiles from simulated halos using dynamical information. This possibility has been discussed before - notably in appendices to \cite{Read_2005} and \cite{Pontzen_2013}, and in \cite{Callingham_2020} - but its possible application to reducing the noise in numerical density estimates has not been explored in detail.  Specifically,  the technique `smears' particles in a snapshot along their orbits, spreading the mass of each across multiple density bins. Such a dynamical approach shares some similarities with certain classical mass modelling techniques \citep[][]{Schwarzschild_1979, Syer_1996} but, unlike these, it does not attempt to match observational constraints to underlying orbits and potentials; rather it constructs these from a simulation snapshot.
The result is a profile which is averaged over a dynamical time, and which consequently has reduced Poisson noise compared to traditional binned estimates at the same resolution.
This, in turn, makes it possible to probe further into the behaviour of the inner regions, at radii where there are very few particles present.

Calculating a density profile through this averaging process inherently assumes an equilibrium, phase-mixed distribution function. This assumption is expected to be significantly broken in the outer parts of a halo approaching the virial radius or beyond. Furthermore, for a practical calculation, we will also assume spherical symmetry (although this assumption could in principle be relaxed).  The gravitational potentials of real and simulated haloes are far from being perfectly spherical. Their shapes tend to be closer to triaxial, especially towards the centre \citep[e.g.][]{Frenk_1988, Jing_2002, Allgood_2006, Orkney_2023}; however it has previously been argued using Hamiltonian perturbation theory that approximating the true triaxial potential by a spherically-averaged version should make little difference to dynamical density estimates if the system is in equilibrium  \citep[][]{Pontzen_2015}. We will return to this point in our discussion.  
Our results focus on the innermost and the outermost regions of haloes to investigate the limits of dynamical halo profiles subject to these coupled assumptions of equilibrium and spherical symmetry.

The rest of the paper is structured as follows. In Section $\S$\ref{methods_section}, we explain the procedure used to generate the dynamical density profiles. In Section $\S$\ref{simulation_section}, we describe the simulation suites and the selection of snapshots analysed in this work. In Section $\S$\ref{results_section}, we present the main results for the dynamical profiles, focusing on the inner and outer regions, and comparing our dynamical technique to traditional binned methods. In Section $\S$\ref{discussion_section}, we discuss the implication of our results and outline possible further work.

\section{Methods} \label{methods_section}

We now describe the methods used to construct dynamical profiles. Section $\S$\ref{create_dyn_densities} considers the construction of a spherically-averaged gravitational potential starting from a simulation snapshot; the calculation of particle orbits within that potential; and finally the computation of the dynamical density profile. In Section $\S$\ref{sec:apo-peri-correction}, we introduce a refinement to the method which improves the accuracy of the orbit integration around apocentre and pericentre. Then, in Section $\S$\ref{potential_iteration_section}, we describe an iterative process via which a self-consistent density--potential pair may be generated.

\subsection{Creating the dynamical density profiles} \label{create_dyn_densities}

We start by assuming that we have a snapshot containing only dark matter particles, centred on the target halo. The spherically-averaged gravitational potential given by all the particles in the snapshot is then calculated in bins of width $\Delta r$ according to the discretized integral 
\begin{equation} \label{potential_eqn}
         \Phi(r_k) = G \sum_{j=1}^{k} \frac{M(<r_j)}{r_j^2} \Delta r,
\end{equation}
where $j$ is an index over the bins, $k$ is the bin number for which the potential is being calculated, and $r_j$ is the radius in the centre of the $j$th bin, taking the value $r_j = (j-1/2) \Delta r$. In addition, $M(<r_j)$ is the mass enclosed within radius $r_j$, and $G$ is the gravitational constant. Although the potential for each bin $k$ is evaluated from quantities at the centre of the bin, the values are assigned to the right edge of the corresponding bins, since $\Phi(r_k)$ represents the average of the potential over the entire bin $k$. The zero point of the potential is set at $r=0$ (the left edge of the first bin).

Equation (\ref{potential_eqn}) is the simplest of several possible choices to perform numerical integration. We tested that adopting a more sophisticated method does not significantly affect the final results.  Therefore, we adopted the simple approach for transparency.

The total number of bins over which $\Phi$ is calculated is determined by the radius of a `bounding sphere' centred around the halo. In addition to choosing the radius at which to truncate the potential, we must also decide how to treat particles whose orbits cross this boundary. In keeping with the core assumption of equilibrium, we make the boundary {\it reflecting}, i.e. particles bounce elastically off it. One may equivalently imagine the potential as having an infinite potential step at the truncation radius. While this is unphysical for any individual particle considered in isolation, across the population it is equivalent to the much more reasonable assumption that the outwards flux through the sphere is balanced by a matching inwards flux. This assumption can be tested by changing the truncation radius; the halo virial radius is a natural first choice, and we will explore the effects of other choices on the final density profile in Section $\S$\ref{bounding}.

Assuming equilibrium, the probability density $p_i(r)$ of finding particle $i$ at radius $r$ is proportional to the time spent by the particle in the infinitesimal interval around that radius: 
\begin{equation} \label{probabs}
p_i(r) = \frac{1}{T_i} \int_{0}^{T_i} \delta (r - r_i(t)) \  \text{d}t \  = \ \frac{2}{T_i} \frac{1}{\dot{r}(r,E_i,j_i,\Phi)},
\end{equation}
where $r_i(t)$ describes the radius as a function of time for the particle (on its spherical idealised orbit), $T_i$ is the period of the orbit, $E_i$ is its specific energy, and $j_i$ is its specific angular momentum. Rather than calculate $T_i$ directly we first calculate an unnormalised version of the probability, $q_{i,k} \equiv (T_i/2) p_i(r_k)$. Here $i$ indexes the particles, and $k$ indexes the spatial bins. By writing the specific energy of a particle as the sum of the potential energy, the kinetic energy due to the angular momentum, and the kinetic energy due to the radial motion, we can solve for $\dot{r}$ and obtain
\begin{equation} \label{probability} 
    q_{i,k} \equiv \frac{1}{\dot{r}(r_k,E_i,j_i,\Phi)} = \left(E_i - \frac{j_i^2}{2r_k^2} - \Phi(r_k)\right)^{-\frac{1}{2}}.
\end{equation}
Note that this expression is only valid between pericentre and apocentre; outside this radial range, it becomes imaginary. However the true probability of finding the particle outside the extrema of its orbit is zero by definition, and therefore one may make Eq.~\eqref{probability} true for all radial bins by taking its real part. We produce a normalized probability for each bin $k$ and particle $i$ according to
\begin{equation}
\begin{aligned}
    p_{i,k} \equiv  \left( \frac{ \mathfrak{Re}\, q_{i,k}}{ \mathfrak{Re}\, \sum_{j} q_{i,j}} \right).
\end{aligned} \label{eq:probability-from-unnomralized}
\end{equation}
where $\mathfrak{Re}$ denotes the real part. If a particle $i$ is on an almost perfectly circular trajectory, it may remain within a single radial bin $k$ for its entire orbit; in this case, the equation above fails and instead a unit probability is assigned to the bin enclosing the original position of the particle in the snapshot, $p_{i,k} = 1$.

\begin{figure}
     \centering
      \includegraphics[width=\columnwidth]{{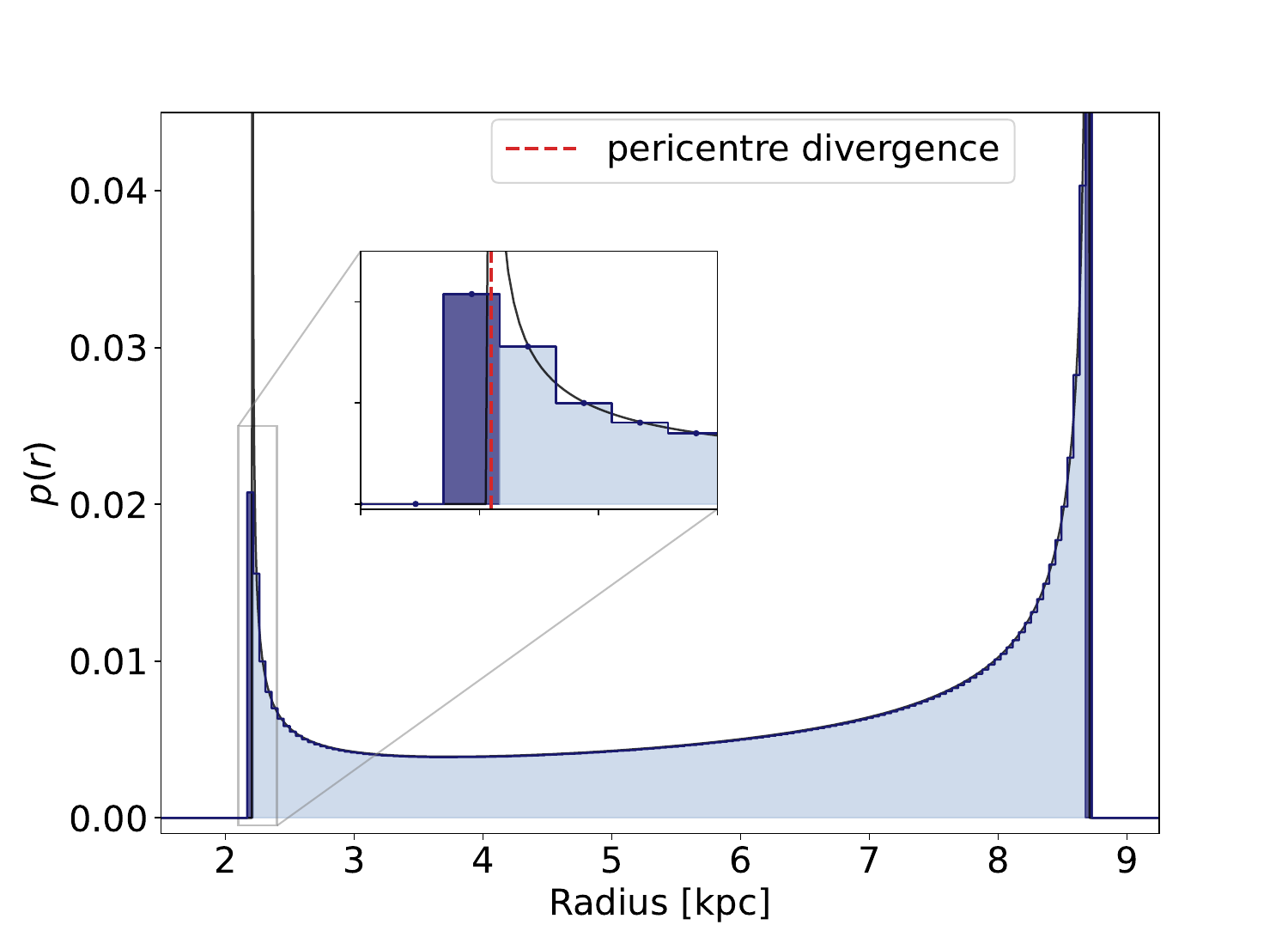}}
     \caption{The binned probability density implied by Eq. (\ref{probability}) evaluated for a typical particle (light-blue bins), with a bin size $\Delta r=\epsilon / 2$, compared with the analytic integrand (black line). The integrand is well behaved for most of the radial range of the orbit, and therefore well approximated by the binned density. However, it has two integrable divergences at pericentre and apocentre (here located at $\sim 2.2$ kpc and $\sim 8.7$ kpc, respectively). Even if the particle never reaches the centre of one of these extremal bins, it may still spend significant time within the bin. Capturing this effect correctly in the binned probability requires the special treatment explained in the text. The dark-blue shaded areas represent the analytical corrections added at the pericentre and apocentre for this orbit.}
     \label{example_orbit}
\end{figure}

The density at the centre of bin $k$ can then be estimated from the set of $p_{i,k}$ as 
\begin{equation}
    \rho(r_k) = \frac{3}{4 \pi} \sum_{i=1}^N \frac{m_i p_{i,k}}{(r_k+\Delta r/2)^3 - (r_k - \Delta r/2)^3 },\label{eq:final-dynamical-denisty}
\end{equation}
where $m_i$ is the mass of each particle $i$, and there are $N$ particles in total.

The statistical errors in the dynamical density profile are estimated using  bootstrapping. For each of $100$ bootstrap samples, we create a mock set of particles by sampling (with replacement) from the actual set of particles in the halo; we then perform the full dynamical density estimate on the mock set of particles.  We determined that 100 bootstrap samples was sufficient to achieve convergence on the 95$\%$ confidence interval; in  Section $\S$\ref{results_section}, our results are shown with these uncertainties as a shaded band.

\subsection{Improving accuracy at apocentre and pericentre}\label{sec:apo-peri-correction}

The function in Eq. (\ref{probability}) has two integrable divergences located at the pericentre and apocentre of each orbit (Figure \ref{example_orbit}). 
Unless the bins are infinitesimally small, the probability of finding the particle in the bin $p_{i,k}$ containing such a divergence might be misestimated. 
To correct for this, in these two bins we use an  approximation scheme based on a local Taylor expansion of the potential. We define the effective potential as $\Phi_{\text{eff}} = \Phi + j^2/(2r^2)$, and expand $\Phi_{\text{eff}}(r_0 +\delta r)$ around $r_0$, where $r_0$ is the divergence point (pericentre or apocentre) for every orbit, i.e., a root of Eq. (\ref{probability}). We now consider the case of a pericentre where the divergence $r_0$ is inside the $k$th bin (i.e., $(k-1) \Delta r < r_0 < k \Delta r$), as an example. The mean value of $\mathfrak{Re} \,\dot r^{-1}$ across the entire bin may be calculated as
\begin{align}
    \bar{q} & \equiv \frac{1}{\Delta r} \int_{r_0}^{r_k} (E_i - \Phi_{\text{eff}}(r))^{-1/2} \text{d}r, \nonumber \\
    & \approx \frac{1}{\Delta r} \int_{r_0}^{r_k} \left( - \frac{\text{d}\Phi_{\text{eff}}}{\text{d}r}\bigg\rvert_{r_0} (r - r_0) \right)^{-1/2} \text{d}r.
\end{align}
Here we have also used the fact that $\Phi_{\text{eff}}(r_0) = E_i$, by definition.
We can furthermore approximate $ {\text{d}\Phi_{\text{eff}}}/{\text{d}r}\rvert_{r_0} \approx {\text{d}\Phi_{\text{eff}}}/{\text{d}r}\rvert_{r_k}$ to avoid having to calculate the exact location of the divergences; this will give us a correction that is accurate to first order. The integration is then analytically tractable, giving
\begin{equation}
    \bar{q} \approx \frac{1}{\Delta r} \frac{2 (E_i - \Phi_{\text{eff}}(r_k))^{1/2}}{\text{d}\Phi_{\text{eff}}(r_k)/\text{d}r|_{r_k}} .
\end{equation}
This analytical estimate of the mean value is then used to represent the value of the probability density function within the pericentre bin $q_k$. 
The apocentre bin is treated in the same way, and both corrections are included before producing the normalized probability according to Eq.~\eqref{eq:probability-from-unnomralized}.

There are two cases in which these corrections cannot be evaluated.
One of them is when an orbit is unresolved (i.e. its probability function only spans one bin), since in that case pericentre and apocentre are coincident.
As previously stated, when this occurs, the particle is given unit probability to be found within the single bin, and corrections are not required.
The apocentre corrections are also ignored when the particle's apocentre falls outside of the radius of the `reflecting wall' which serves as the boundary for the halo. Since the particles can be thought of as being reflected back once they hit the boundary, their radial paths are truncated at the location of the wall, and no apocentre corrections are required.

\subsection{Iterating the potential} \label{potential_iteration_section}

The dynamical density profile given by Eq.~\eqref{eq:final-dynamical-denisty} implies also a mass profile $M(<r)$ and therefore a potential $\Phi(r)$ through Eq.~\eqref{potential_eqn}. However, the potential used in producing the density estimate was initialized directly using the particle radii from the original snapshot. The overall procedure, therefore, results in an inconsistent potential--density pair. 
The difference between the mass distribution is especially evident in the inner regions because our potential is calculated without softening, and the pericentres of orbits can therefore reach radii closer to the centre of the haloes. 
To resolve this discrepancy, we iterate until a self-consistent density-potential configuration pair is reached. Over the course of the iterations, the gravitational potential from the simulation is gradually transformed into the potential inferred from the dynamical density profile.
This technique also removes any discontinuities in the derivatives of the potential at small radii due to the finite particle number. 
    
The iteration process involves a series of steps:
\begin{enumerate}
    \item A dynamical density profile is first obtained as described in Sections $\S$\ref{create_dyn_densities} and $\S$\ref{sec:apo-peri-correction}.
    \item The mass distribution implied by the dynamical profile is calculated according to 
    \begin{equation}
    \begin{aligned}
        M(<r_j + \Delta r/2) = \sum_{i=1}^{N} \sum_{k=1}^{j} m_i p_{i,k}.
    \end{aligned}
    \end{equation} 
    The mass at the centre of the bin ($r_j$) is then obtained by averaging the mass at adjacent edges.
    \item The new mass distribution is inserted into Eq. (\ref{potential_eqn}) to evaluate a new gravitational potential.
    \item The angular momenta of the particles is assumed to be unchanged, and the energies are updated by keeping the radial action constant at first order (see below).
    \item The cycle is repeated, starting from point (ii) and using the updated dynamical profile, until convergence in the dynamical profile is reached.
\end{enumerate}

\noindent Evolving the gravitational potential into the new configuration will affect the phase space distribution of the particles. Hence we require the energies of the particles to change accordingly. In step (iv) the updated energies are calculated by keeping the radial action of each particle constant to first order, 
\begin{equation} \label{constant_radial_action}
    J_r(E_{\mathrm{new},i}, j_i, \Phi_{\mathrm{new}}) = J_r(E_{\mathrm{old},i}, j_i, \Phi_\mathrm{old}) + \mathcal{O}(\Delta \Phi^2),
\end{equation}
\noindent for each particle $i$, where $E_{\mathrm{new},i}, \Phi_\mathrm{new}$ and $E_{\mathrm{old},i}, \Phi_\mathrm{old}$ are the specific energy and the potential after and before the iteration respectively, and $\Delta \Phi = \Phi_{\mathrm{new}} - \Phi_{\mathrm{old}}$. 
We keep $J_r$ constant since we can interpret each iteration as a small change to the potential of the halo, akin to an adiabatic relaxation. This process does not correspond to a literal physical evolution of the halo in time, but an adiabatic transformation is nonetheless the most conservative way to map orbits from the potential at each iteration to the next. In other words, we assume that the action distribution of the particles in the simulation is sampled from an underlying `true’ distribution (as would be attained by a simulation of infinite resolution). We then  recover the potential implied by the dynamical profile given this action distribution. If we assume that the change to the potential between iterations is sufficiently small, we only need update the actions at first order in the potential change, i.e. Eq.(\ref{constant_radial_action}).

The definition of the radial action is 
\begin{equation}
J_r(E,j,\Phi) = \frac{2}{\pi}\int_{r_{\mathrm{peri}}}^{r_{\mathrm{apo}}} \sqrt{E-\frac{j^2}{2 r^2} - \Phi(r)}\, \mathrm{d}r\,.
\end{equation}
With this in hand, we solve Eq. (\ref{constant_radial_action}) to first order in the quantities $\Delta \Phi$ and $\Delta E_i = E_{\mathrm{new},i} - E_{\mathrm{old},i}$. By Taylor expanding, we find
\begin{equation}
    \Delta E_i \approx \frac{\int^{r_{\text{apo}}}_{r_{\text{peri}}} \Delta \Phi(r) \left(E_{\mathrm{old},i} - \frac{j_i^2}{2r^2} - \Phi_\mathrm{old}(r)\right)^{-1/2} \text{d}r} { \int^{r_{\text{apo}}}_{r_{\text{peri}}} \left(E_{\mathrm{old},i} - \frac{j_i^2}{2r^2} - \Phi_\mathrm{old}(r)\right)^{-1/2} \text{d}r } = \langle \Delta \Phi \rangle,
\end{equation}
i.e. the change in energy is equal to the average of the change in potential, weighted by the probability of finding the particle at a given radius. (At first order, the changes to the values of apocentre and pericentre of the orbit do not contribute to $\Delta E$, and can therefore be neglected.) 

The first iteration produces a significant change in the inner density distribution but after approximately 3 iterations, convergence in the dynamical profile is reached (i.e. the changes in the density profiles become significantly smaller than the bootstrap-determined uncertainties). We will discuss this further in Section $\S$\ref{sec:effect-of-potential-iterations} below.

\section{The simulation snapshots} \label{simulation_section}

We analyse a selection of seven snapshots drawn from cosmological zoom simulations of dark matter haloes spanning a wide range of masses, from ${\sim}10^9 \text{M}\textsubscript{\(\odot\)}$ to ${\sim}10^{12} \text{M}\textsubscript{\(\odot\)}$ (see Table \ref{table_snapshots}).

\begin{table*}
	\centering 
 \caption{Properties (softening length, particle mass, number of particles, virial radius, virial mass, and brief comments on the density structure) of the seven haloes investigated in this work. The haloes can be grouped into 3 main categories based on their virial mass, from dwarf to Milky Way mass. The number of particles refers to the particles enclosed by each halo's virial radius at $z=0$. \label{table_snapshots}}
  \setlength{\tabcolsep}{5pt}
	\begin{tabular}{ccccccccc}
		\hline
		\textbf{Halo} & \textbf{Figure} & \textbf{Resolution} & \textbf{$\epsilon$ (kpc)} & \textbf{Particle Mass ($\boldsymbol{\text{M}\textsubscript{\(\odot\)}}$)} & \textbf{Number of Particles} & \textbf{$r_{\text{vir}}$ (kpc)} & \textbf{Virial Mass ($\boldsymbol{\text{M}\textsubscript{\(\odot\)}}$)} & \textbf{Structure} \rule{0mm}{4mm} \\[1mm]  
		\hline
\rule{0mm}{4mm}

\multirow{2}{*}{1445} & \multirow{2}{*}{\ref{all_1400s_halos}, top} & \multicolumn{1}{c }{Low} & \multicolumn{1}{c }{0.095} & \multicolumn{1}{c }{$7.1\times10^4$} & \multicolumn{1}{c }{$3\times10^4$} & \multicolumn{1}{c }{41.7} & & \\ 
                                 & \multicolumn{1}{c }{} & \multicolumn{1}{c }{High} & \multicolumn{1}{c }{0.012} & \multicolumn{1}{c }{$1.1\times10^3$} & \multicolumn{1}{c }{$2\times10^6$} & \multicolumn{1}{c }{41.5} & \multirow{-2}{*}{$2 \times 10^9$} & \multirow{-2}{*}{\makecell{Substructures at large $r$; \\ dynamical equilibrium}} \vspace{0.2cm}\\ 

\multirow{3}{*}{1459} & \multirow{3}{*}{\ref{all_1400s_halos}, bottom} & \multicolumn{1}{c }{Low} & \multicolumn{1}{c }{0.095} & \multicolumn{1}{c }{$7.1\times10^4$} & \multicolumn{1}{c }{$3\times10^4$} & \multicolumn{1}{c }{41.4} & \multirow{3}{*}{$2 \times 10^9$} & \multirow{3}{*}{\makecell{Substructures at large $r$; \\ dynamical equilibrium}}\\ 
                                 & \multicolumn{1}{c }{} & \multicolumn{1}{c }{High} & \multicolumn{1}{c }{0.012} & \multicolumn{1}{c }{$1.1\times10^3$} & \multicolumn{1}{c }{$2\times10^6$} & \multicolumn{1}{c }{41.1} & \\ 
& \multicolumn{1}{c }{} & \multicolumn{1}{c }{Ultra-high} & \multicolumn{1}{c }{0.006} & \multicolumn{1}{c }{$1.4\times10^2$} & \multicolumn{1}{c }{$1\times10^7$} & \multicolumn{1}{c }{41.1} &  \vspace{0.2cm}\\

\multirow{3}{*}{600} & \multirow{3}{*}{\ref{all_600s_halos}, top} & \multicolumn{1}{c }{Low} & \multicolumn{1}{c }{0.095} & \multicolumn{1}{c }{$7.1\times10^4$} & \multicolumn{1}{c }{$8\times10^4$} & \multicolumn{1}{c }{56.8} & \multirow{3}{*}{$5 \times 10^9$} & \multirow{3}{*}{\makecell{Low res: recent merger, \\ disequilibrium (cusp). \\ Higher res: equilibrium}} \\ 
                                 & \multicolumn{1}{c }{} & \multicolumn{1}{c }{High} & \multicolumn{1}{c }{0.012} & \multicolumn{1}{c }{$1.1\times10^3$} & \multicolumn{1}{c }{$5\times10^6$} & \multicolumn{1}{c }{56.2} & & \\ 
& \multicolumn{1}{c }{} & \multicolumn{1}{c }{Ultra-high} & \multicolumn{1}{c }{0.006} & \multicolumn{1}{c }{$1.4\times10^2$} & \multicolumn{1}{c }{$4\times10^7$} & \multicolumn{1}{c }{56.2} &  \vspace{0.2cm}\\

\multirow{2}{*}{605} & \multirow{2}{*}{\ref{all_600s_halos}, middle} & \multicolumn{1}{c }{Low} & \multicolumn{1}{c }{0.095} & \multicolumn{1}{c }{$7.1\times10^4$} & \multicolumn{1}{c }{$7\times10^4$} & \multicolumn{1}{c }{55.0} & \multirow{2}{*}{} & \\ 
                                 & \multicolumn{1}{c }{} & \multicolumn{1}{c }{High} & \multicolumn{1}{c }{0.012} & \multicolumn{1}{c }{$1.1\times10^3$} & \multicolumn{1}{c }{$4\times10^6$} & \multicolumn{1}{c }{54.7} & \multirow{-2}{*}{$5 \times 10^9$} & \multirow{-2}{*}{\makecell{Minimal substructure; \\ dynamical equilibrium}} \vspace{0.2cm}\\

\multirow{2}{*}{624} & \multirow{2}{*}{\ref{all_600s_halos}, bottom} & \multicolumn{1}{c }{Low} & \multicolumn{1}{c }{0.095} & \multicolumn{1}{c }{$7.1\times10^4$} & \multicolumn{1}{c }{$7\times10^4$} & \multicolumn{1}{c }{56.3} & \multirow{2}{*}{} & \\ 
                                 & \multicolumn{1}{c }{} & \multicolumn{1}{c }{High} & \multicolumn{1}{c }{0.012} & \multicolumn{1}{c }{$1.1\times10^3$} & \multicolumn{1}{c }{$5\times10^6$} & \multicolumn{1}{c }{56.2} & \multirow{-2}{*}{$5 \times 10^9$} & \multirow{-2}{*}{\makecell{Pre-merger; \\ significant disequilibrium}} \vspace{0.2cm}\\

\multirow{2}{*}{685} & \multirow{2}{*}{\ref{all_milkyway_halos}, top} & \multicolumn{1}{c }{Low} & \multicolumn{1}{c }{0.142} & \multicolumn{1}{c }{$2.3\times10^5$} & \multicolumn{1}{c }{$5\times10^6$} & \multicolumn{1}{c }{349.0} & \multirow{2}{*}{} & \\ 
                                 & \multicolumn{1}{c }{} & \multicolumn{1}{c }{High} & \multicolumn{1}{c }{0.035} & \multicolumn{1}{c }{$2.9\times10^4$} & \multicolumn{1}{c }{$4\times10^7$} & \multicolumn{1}{c }{346.9} & \multirow{-2}{*}{$1 \times 10^{12}$} & \multirow{-2}{*}{\makecell{Minimal substructure; \\ dynamical equilibrium}} \vspace{0.2cm}\\

\multirow{2}{*}{715} & \multirow{2}{*}{\ref{all_milkyway_halos}, bottom} & \multicolumn{1}{c }{Low} & \multicolumn{1}{c }{0.142} & \multicolumn{1}{c }{$2.3\times10^5$} & \multicolumn{1}{c }{$6\times10^6$} & \multicolumn{1}{c }{358.4} & \multirow{2}{*}{} & \\ 
                                 & \multicolumn{1}{c }{} & \multicolumn{1}{c }{High} & \multicolumn{1}{c }{0.035} & \multicolumn{1}{c }{$2.9\times10^4$} & \multicolumn{1}{c }{$5\times10^7$} & \multicolumn{1}{c }{357.3} & \multirow{-2}{*}{$1 \times 10^{12}$} & \multirow{-2}{*}{\makecell{Minimal substructure; \\ dynamical equilibrium}} \\[1mm] 
    
		\hline
	\end{tabular}
\end{table*}

The five smallest haloes are part of the Engineering Dwarfs at Galaxy Formation’s Edge (EDGE) project \citep[][]{Agertz_2019, Rey_2019, Rey_2020, Orkney_2021}; the two largest haloes were taken from the \textsc{vintergatan-gm} project, which in turn uses the initial conditions described by \citet[][]{Rey_2021_vintergatan}. 
Both suites of simulations assume a $\Lambda$CDM cosmology: EDGE adopts cosmological parameters based on data from \citet[][]{Plank_2014} ($ \Omega_m = 0.309 $, $ \Omega_{\Lambda} = 0.691 $, $H_0 = 67.77$ km $\text{s}^{-1} \text{Mpc}^{-1} $) with a box size of 50 Mpc, while \textsc{vintergatan-gm} uses cosmological parameters from \citet[][]{Plank_2016} ($\Omega_m = 0.314$, $\Omega_{\Lambda} = 0.686$, $H_0 = 67.27$ km $\text{s}^{-1} \text{Mpc}^{-1} $) with a box size of 73 Mpc.
As previously stated, we consider the dark-matter-only simulations from these suites, i.e. they do not contain any baryonic components; hence steep cusps are expected in the central regions of the density profiles.

The selected haloes were re-simulated at two different resolutions; the particle mass ratio between the lower and in the higher resolution runs is 64 (for EDGE) and 8 (for \textsc{vintergatan-gm}). 
Both suites of simulations are generated using the adaptive mesh refinement (AMR) code \textsc{ramses} \citep[][]{Teyssier_2002}. The mesh is refined whenever a grid cell contains more than 8 particles; consequently, the softening lengths are adaptive and we provide a softening scale estimate $\epsilon$ equal to the size of the smallest grid cell used for gravity calculations. 
We call \textit{low resolution} the simulations with softening scale of 0.095 kpc (0.142 kpc for the \textsc{vintergatan-gm} haloes), and \textit{high resolution} the ones with softening of 0.012 kpc (0.035 kpc for the \textsc{vintergatan-gm} haloes). \textit{Ultra-high resolution} runs with softening scale $\sim 0.006$ kpc are also available for some EDGE simulations. 
All the snapshots analysed in the current work are taken at the present day ($z=0$).

Simulation snapshots are loaded using pynbody \citep{pynbody}. Before processing, each halo is centred using the shrinking-sphere method of \cite{Power_2003}; the central 1 kpc is used to calculate a centre of mass velocity, which is then subtracted from all particles. We also calculate a virial radius, $r_{\mathrm{vir}}$, defined to be the radius at which the enclosed mean density is equal to 178 times the cosmic mean. 

All particles interior to the reflecting wall at the time of the snapshot are included in the calculations. 
Some of the selected haloes contain large substructures, especially in their outskirts; these are deliberately retained in our analysis in order to test the limits of the assumption of equilibrium. 
The reflecting boundary described in Section $\S$\ref{create_dyn_densities} was placed at 120 kpc for the haloes with mass $ \lesssim 5
\times 10^9 \text{M}\textsubscript{\(\odot\)}$. This is between 2 and 3 times the size of their virial radii, a choice which allows us to explore how the dynamical information affects the density distribution in their outer regions. The boundary for the two largest haloes was placed at 350 kpc, which is approximately the location of their virial radii, and was not extended to larger radii in this work because the `zoomed' region of these haloes is only twice the virial radius, beyond which low resolution particles are present. 
For efficiency, the dynamical profiles of the two largest haloes are generated using only a randomly selected fraction (a third) of the particles.

While it is not possible to recreate precisely the in-simulation softening $\epsilon$ with a spherical approximation, it is clear that the bin width $\Delta r$ must be comparable to $\epsilon$ in order that the potential is meaningful. We found that our results were insensitive to the precise bin width chosen, provided that it is of this order, and therefore chose to fix  $\Delta r = \epsilon / 2$. This choice of bin width is sufficiently small to allow investigation of the dynamically-inferred density profile close to the halo centre. We note that for $r \lesssim 3 \epsilon \equiv r_{\text{conv}}$ the effect of spurious relaxation in simulation becomes important and a profile constructed through direct particle binning is poorly resolved. Detailed studies of convergence \citep[e.g.][]{Power_2003, Gao_2012, Ludlow_2019} show that the value of $r_{\text{conv}}$ must be determined empirically for each simulation setup, and any relation to the softening length $\epsilon$ is approximate; the scale is mainly dictated by the number of particles present in the innermost regions. Our comparisons of binned profiles between high and low resolution simulations below confirm that $r_{\text{conv}} \sim 3\epsilon$ gives a sufficiently good approximation to the innermost reliable radius of the low resolution binned profiles\footnote{Using Eq.(14) in \cite{Ludlow_2019} to calculate the convergence radius, with the constant $\kappa_{P03}$ calibrated using a subset of our haloes at different resolutions, we obtained estimates between $2.6\epsilon$ and $4.5\epsilon$ across all snapshots. Our $r_{\text{conv}} = 3\epsilon$ estimate, therefore, falls within this range. }.

\begin{figure*}
     \centering
         \begin{subfigure}[b]{\textwidth}
         \centering
         \includegraphics[width=0.8\textwidth]{{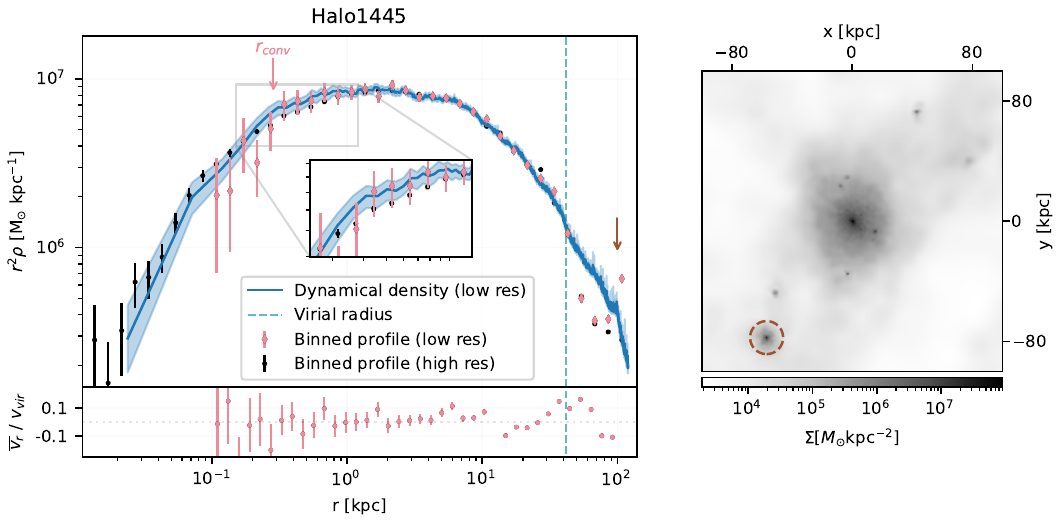}}
     \end{subfigure}
        \begin{subfigure}[b]{\textwidth}
        \hspace{1.8cm}\includegraphics[width=0.8\textwidth]{{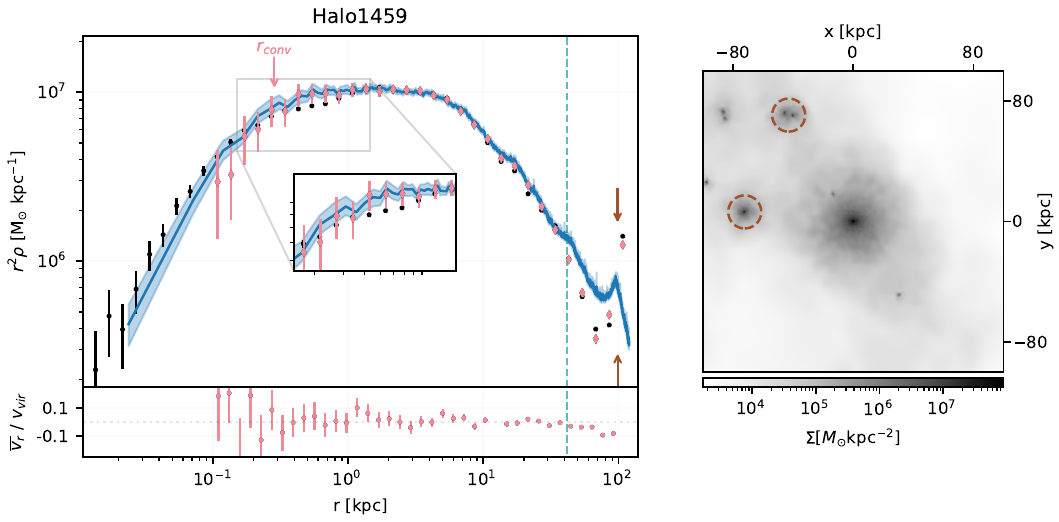}}
     \end{subfigure}
     \caption{Density profiles multiplied by $r^2$ (left) and images of the dark matter density projected down the $z$ axis (right) for our two lowest-mass dwarf haloes ($M \sim 2 \times 10^9 \text{M}\textsubscript{\(\odot\)}$).
     The dynamical density profiles obtained from the low resolution snapshots (blue lines) agree very well with both the low and high resolution binned profiles (pink and black points) for most of the radial extent of all the haloes. The largest variations between the dynamical and binned estimates are observed in the outer regions, beyond the virial radius, where large substructures in the outskirts cause spikes in the mass distribution.
     Any such substructures with mass greater than 3$\%$ of the mass of the main halo are shown by brown circles in the halo images, and by corresponding brown arrows in the dynamical profile plots. The panels below the density profiles show the variations in the average radial velocity of the particles contained within concentric shells as a fraction of the virial velocity, which can be used to quantify how close the low resolution halo is to equilibrium. The pink arrows indicate the radius corresponding to 3 times the value of the softening scale of the low resolution simulations (i.e. $r_{\text{conv}}$ for the low resolution binned profiles). }
     \label{all_1400s_halos}
\end{figure*}

\begin{figure*}
     \centering
         \begin{subfigure}[b]{\textwidth}
         \centering
         \includegraphics[width=0.8\textwidth]{{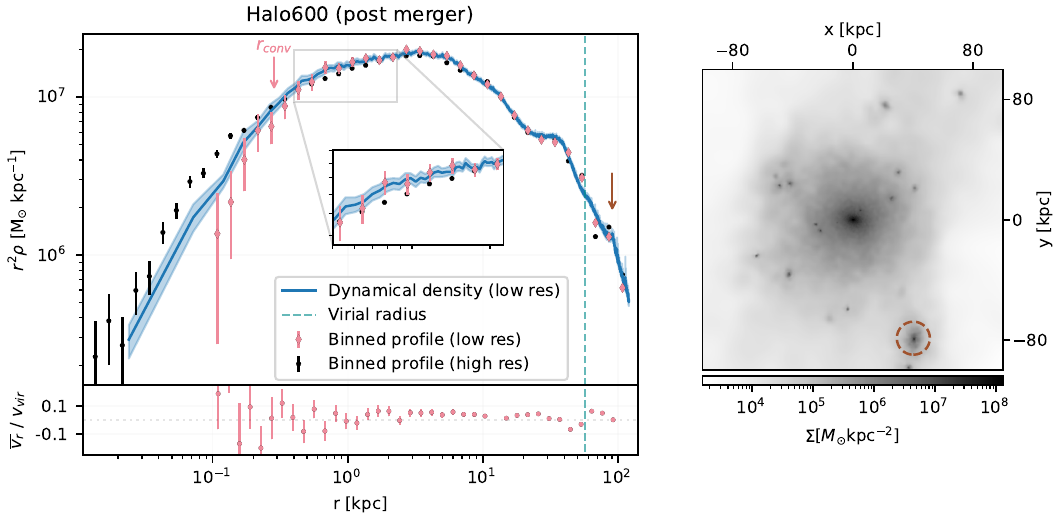}}
     \end{subfigure}
        \begin{subfigure}[b]{\textwidth}
        \hspace{1.8cm}\includegraphics[width=0.8\textwidth]{{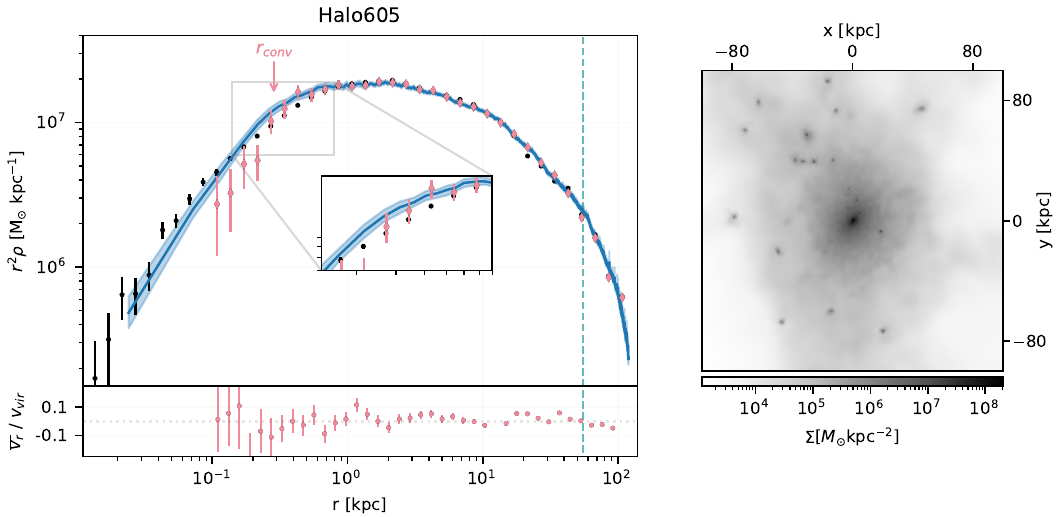}}
     \end{subfigure}
       \begin{subfigure}[b]{\textwidth}
        \hspace{1.8cm}\includegraphics[width=0.8\textwidth]{{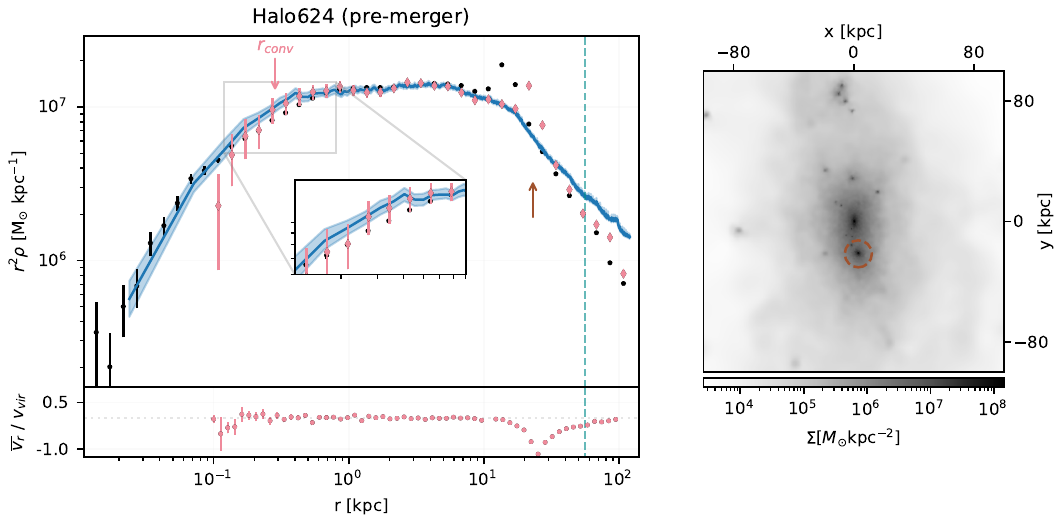}}
     \end{subfigure}
     \caption{Same as Figure \ref{all_1400s_halos} but for the three intermediate-mass dwarf haloes ($M \sim 5 \times 10^9 \text{M}\textsubscript{\(\odot\)}$). Similarly to the other cases, the dynamical density profiles from the low resolution snapshots agree well with both binned profiles. Halo600 is an outlier since it recently had a merger close to the halo's centre which disrupted the equilibrium in the inner regions; as a result the plot of $\bar{\text{v}}_r / \text{v}_{\text{vir}}$ shows significant deviations from zero at small radii. Halo624 has a large substructure within its virial radius which will reach the centre of the main halo and merge with it in the next $\sim 500$ Myrs. (The structure is found slightly closer to the centre in the high resolution simulation.) The significant disruption caused by this substructure to the halo's equilibrium is also evident in the average radial velocity panel, but our dynamical method nonetheless recovers a sensible `smoothed' density profile.}
     \label{all_600s_halos}
\end{figure*}

\begin{figure*}
     \centering
         \begin{subfigure}[b]{\textwidth}
         \centering
         \includegraphics[width=0.8\textwidth]{{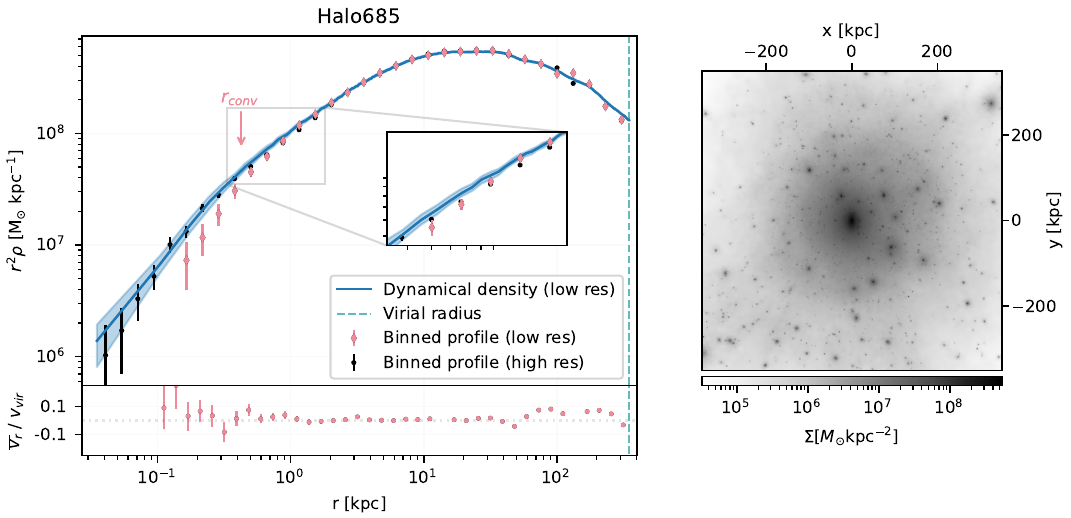}}
     \end{subfigure}
        \begin{subfigure}[b]{\textwidth}
        \hspace{1.8cm}\includegraphics[width=0.8\textwidth]{{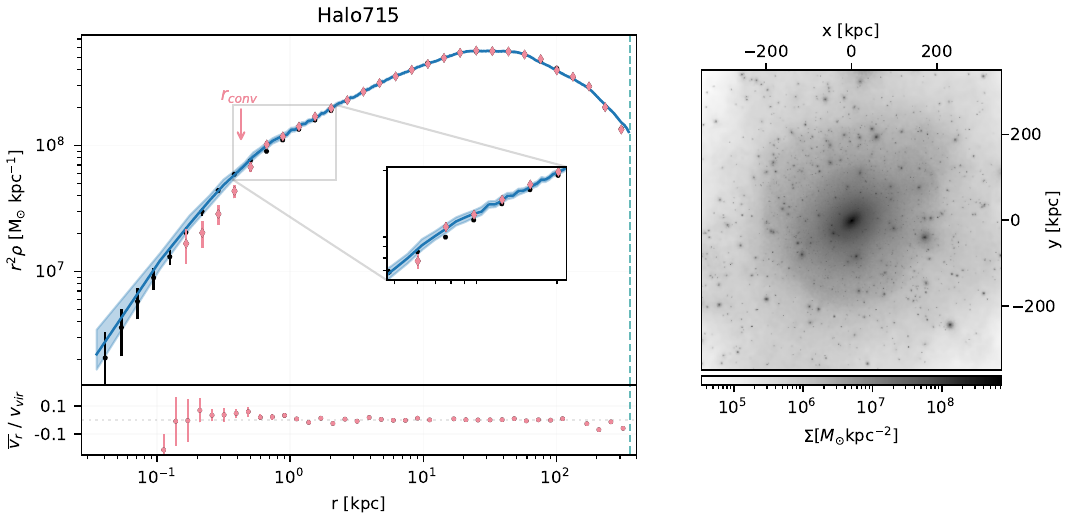}}
     \end{subfigure}
     \caption{Same as Figures \ref{all_1400s_halos} and \ref{all_600s_halos} but for the two most massive ($M \sim 10^{12} \text{M}\textsubscript{\(\odot\)}$) out of all seven haloes. Similarly to the other haloes, the dynamical density profiles from the low resolution snapshots agree well with both the low and high resolution binned profiles. For efficiency, the dynamical profiles for these haloes were generated using only a randomly selected fraction (a third) of all the particles within the halo and therefore even smaller errors on the dynamical density profile are achievable in principle. In these examples, all substructures are small (less than $1\%$ of the halo mass) and do not have a visible effect on the density profiles.}
     \label{all_milkyway_halos}
\end{figure*}

\section{Results} \label{results_section}

In this Section, we present and discuss the dynamical density profiles of our dark matter haloes. In each case, we calculate dynamical profiles from the low resolution snapshots and compare them with binned profiles from both low resolution and high resolution snapshots. The profiles are shown in Figures \ref{all_1400s_halos}, \ref{all_600s_halos}~and~\ref{all_milkyway_halos} (for lowest-mass dwarf, intermediate-mass dwarf and Milky-Way-mass haloes respectively), alongside images of the haloes' dark matter density projected down the $z$ axis.
We compare our dynamical profiles (blue lines) to the traditional binned estimates from both the high and the low resolution snapshots (black and pink points respectively), which are plotted down to their estimated softening length (see Table~\ref{table_snapshots}). Inset panels show the inner density profile in greater detail. 

Overall, the dynamical profiles (blue lines), obtained from the low resolution simulations, agree well with the low resolution binned profiles (pink points) for the majority of the radial extent of the haloes. 
The $95\%$ bootstrap-determined uncertainties on the dynamical profiles are shown as shaded blue bands, and are significantly smaller than the $95\%$ Poisson noise on direct binned estimates at the same resolution (pink error-bars). This follows from the fact that the particles in the original snapshot are now spread across multiple density bins, hence providing better statistics.

By dividing the total volume occupied by each halo into thin shells, we can also calculate the average radial velocities of the particles contained within the shells. These are shown for the low resolution simulations in the panels below the density profiles in Figures \ref{all_1400s_halos} -- \ref{all_milkyway_halos}. These values will help us discuss below how well the assumption of equilibrium holds for each halo.

We will first discuss the behaviour of the dynamical profiles in the inner regions (around or even interior to the traditional convergence radius; Section $\S$\ref{sec:inner-regions}), then in the outer regions (around and beyond the virial radius; Section $\S$\ref{sec:outer-regions}).

\subsection{Inner regions}\label{sec:inner-regions}

The direct comparison of dynamical profiles (blue lines) with binned profiles from higher resolution simulations (black points) is of considerable interest: it addresses the question of whether our technique can partially correct for finite particle number in the innermost regions of the halo.

At radii below the approximate convergence radius of the low resolution binned profiles ($r_{\mathrm{conv}} = 3 \epsilon$, indicated by the pink arrows in Figures \ref{all_1400s_halos}, \ref{all_600s_halos}, and \ref{all_milkyway_halos}), our dynamical density cusps are steeper than the traditional binned profiles at the same resolution. This is particularly clear in the case of the Milky-Way-mass halos (Figure~\ref{all_milkyway_halos}).
Comparing our results to the binned distribution of the high resolution simulations (black points), we see that the dynamical method is, in nearly all cases, able to predict the `cuspier' behaviour of higher resolution simulations below $r_{\text{conv}}$. This is especially evident in the larger haloes due to the smaller Poisson noise in the central regions. Halo600 is an exception in which the dynamically predicted density is substantially lower than that in the high resolution simulation;  Section $\S$\ref{sec:challenge-differing-resolutions} considers that case in some detail, and more broadly discusses caveats about making comparisons between low and high resolution simulations. Nonetheless, in the other cases studied, the dynamically predicted cusp extends below $r_{\text{conv}}$ of the low resolution simulations, where very few particles are present at the time of the snapshot\footnote{For the EDGE haloes there are only 150-250 particles below $r_{\text{conv}}$, and 350-450 for the \textsc{vintergatan-gm} haloes; this is $0.38\%$ and $0.007\%$ of the number of particles enclosed by the virial radius of the two simulation suites, respectively.}. As well as being less biased than the binned profiles, our dynamical profiles also have lower numerical noise. On average across all halos, the uncertainties at small radii (between $\epsilon$ and $r_{\text{conv}}$) are reduced by a factor of 12 compared to traditional binned estimates. Thus, our technique uses information about the entire phase-space of the particles to produce more precise central density profiles which partially correct for the effects of softening and which are less subject to Poisson noise.

\begin{figure}
	\includegraphics[width=\columnwidth]{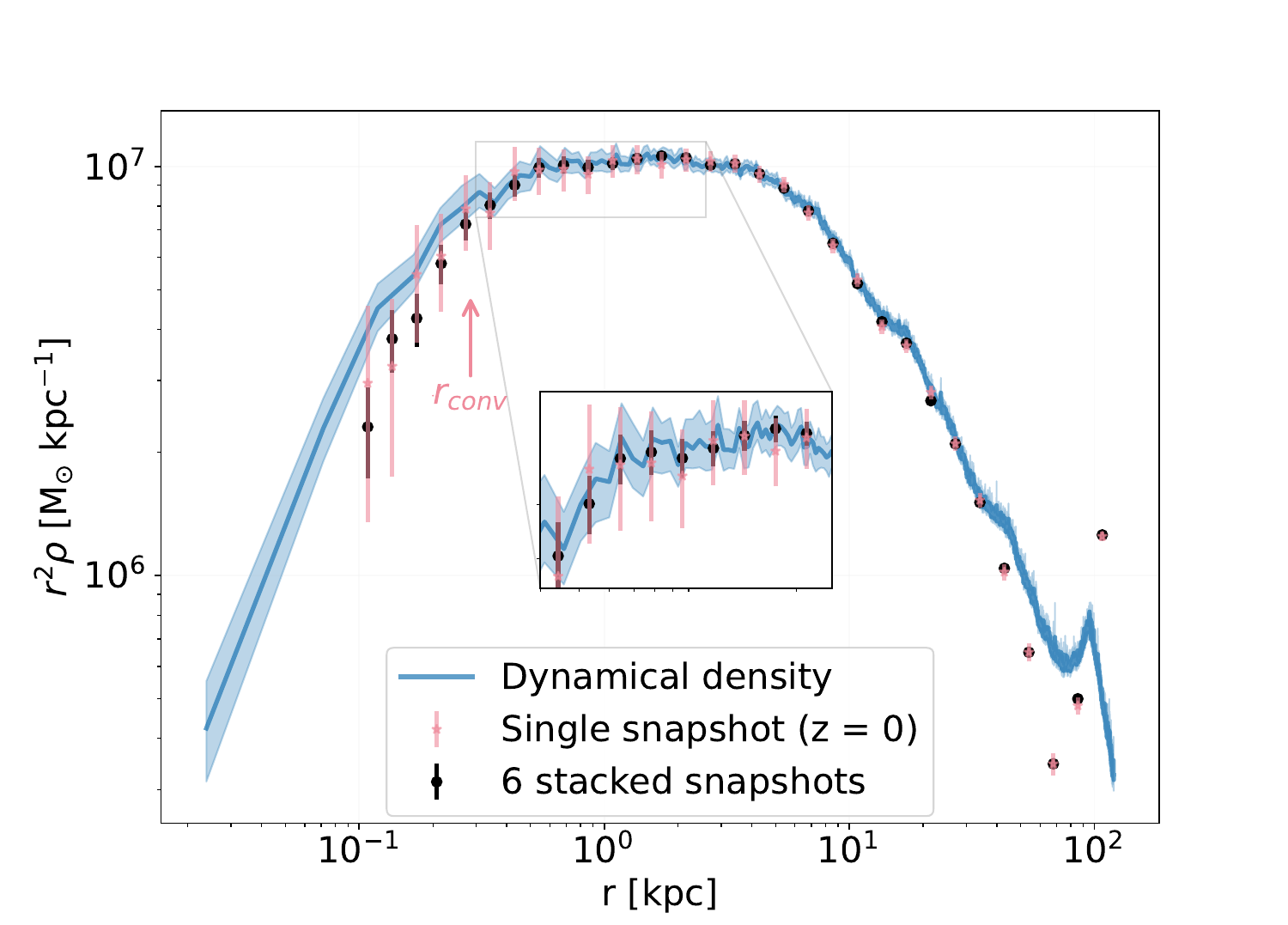}
    \caption{Binned density profile multiplied by $r^2$ obtained by stacking 6 consecutive snapshots from the low resolution simulation of Halo1459 (black points). The dynamical density profile (blue line) and the binned profile obtained from a single snapshot at $z=0$ (pink points) are also shown for comparison. Although Poisson noise is mitigated in the stacked profile, the method cannot correct the systematic softening and relaxation errors, and therefore underestimates central densities, unlike the dynamical profile.}
    \label{stacked_profiles}
\end{figure}

Poisson noise could also be mitigated by stacking binned profiles from adjacent snapshots (similarly to the procedure outlined in \citealt[][]{Vasiliev_2014}{}{}{}). Figure \ref{stacked_profiles} shows an example of the binned profile obtained by stacking 6 adjacent snapshots of Halo1459. This is compared to our dynamical density profile (blue line) and to the binned profile obtained from a single snapshot at $z=0$ (pink points). Stacking the profiles results in considerable reduction in shot noise, similar to the effect observed in the dynamical profile. 
However, the method fails to reproduce the steeper central gradient observed in the dynamical profile below the convergence radius, which implies a significant disagreement with the binned profile from the high resolution runs.
This is due to the fact that the stacked profile retains the effects of gravitational softening and of relaxation caused by encounters between the particles in the low resolution simulation as it evolves over time. By contrast, in the dynamical profile no softening is used and the orbits are integrated independently of each other, allowing the iteration process (Section $\S$\ref{potential_iteration_section}) to correctly recover a steep central cusp. 
While in principle the stacked profile could also be iterated by combining it with our dynamical method, this would entail significant complexity due to the starting potentials in each snapshot differing from each other, as well as from the final combined potential. We therefore leave any investigation of such a combined stacked-dynamical profile to future work.

At radii just larger than $r_{\text{conv}}$, we notice a small but statistically significant density excess in both the binned and dynamical low resolution profiles when compared with the high resolution binned profiles. This excess only covers a few density bins and is more evident for some haloes (e.g. Halo605 and 624) than others; see the inset panels zoomed in on this radius in Figure \ref{all_600s_halos}. 
Since this feature is also present when using binned methods, it must be unrelated to the inclusion of dynamical information into the calculations. We therefore leave investigation to a future study.

\subsubsection{The challenge of direct comparisons between differing resolutions}\label{sec:challenge-differing-resolutions}

Overall, the improvement offered by dynamical profiles over binned profiles  is significant: the uncertainties at small radii are significantly mitigated compared to binned estimates, making it a substantially more precise technique. Qualitatively, it is clear that the dynamical profiles reproduce steeper profiles which appear to be in agreement with higher resolution simulations within the 95\% error bounds. 
However, quantifying how accurate the dynamical estimates are compared to the true density distributions (i.e. the density profiles that would be obtained from simulations of infinite resolution) is difficult for two reasons. The first is the problem of formulating a suitable comparison summary statistic; the second is the impact of small differences in halo formation and merger history on the final profile. We will describe each of these in turn.

The most natural way to measure the accuracy of a low resolution density profile would be to construct a chi-squared test to decide whether the binned or dynamical profiles more accurately predict the high resolution result. However, the size of the statistical errors on the dynamical profile are substantially smaller than those on the binned profile, putting the dynamical profiles at an automatic disadvantage in such a test. Even if one were to artificially inflate the dynamical profile error estimates, the results would remain very sensitive to the precise radial range over which the statistic is calculated. The dynamical profiles clearly predict more accurate densities interior to $r_{\mathrm{conv}}$, but outside this radius the situation is more nuanced. In particular, at large radii, the dynamical profiles' tendency to wash out substructure would lead to a heavy $\chi^2$ penalty (as will be discussed in Section $\S$\ref{sec:outer-regions} below). There is therefore no straight-forward quantitative measurement of the improvement offered by dynamical density profiles, despite the clear qualitative advantages in the cusp region.

The second challenge relates to recent events in the formation and merger history, and is most clearly seen in the case of Halo600 (shown at the top of Figure \ref{all_600s_halos}). As with the other examples, the gradient of the dynamical profile interior to $r_{\text{conv}}$ is steeper than the low resolution binned profile; however, unlike the other cases, the steepening in Halo600 is insufficient to reach agreement with the high resolution binned profile.
The reason can be traced to the halo's recent history in the respective simulations. 
The low resolution version of Halo600 underwent a minor merger at $z=0.03$ ($\sim 70$ Myrs before present day). This merger only occurred in the low resolution version of the simulation. Although the mass of the merger is relatively small (${\sim}10^8 \text{M}\textsubscript{\(\odot\)}$, around $2\%$ of the total host mass), its centre of mass before disruption is located within 1 kpc of the centre of mass of the main halo. 
By tracking the particles that formed the subhalo to $z=0$, we find that they have traversed the halo from one side to the other, and remain in disequilibrium. The out-of-equilibrium behaviour is also visible as large fluctuations in the binned radial velocities as seen in the lower panel of the Halo600 plot in Figure~\ref{all_600s_halos}. Despite this, note that the dynamical density profile still performs somewhat better than the binned profile.

From the above analysis, we deduce that even a relatively small merger might affect the equilibrium of a halo. 
A statistical study on a larger sample of haloes is necessary to constrain the exact relationship between merger-to-main halo mass ratio and the effect that the merger events have on the dynamical profile. Other features will also play a role, such as the object infall velocity or the angle of collision.
The investigation of these effects is beyond the scope of this work.

\subsubsection{Effect of potential iterations}\label{sec:effect-of-potential-iterations}

Having established that dynamical profiles offer an accuracy improvement over binned profiles near the centres of halos, albeit one that is hard to quantify, we now consider the effect of the iterative part of our algorithm (Section $\S$\ref{potential_iteration_section}) in achieving this. 

\begin{figure}
	\includegraphics[width=\columnwidth]{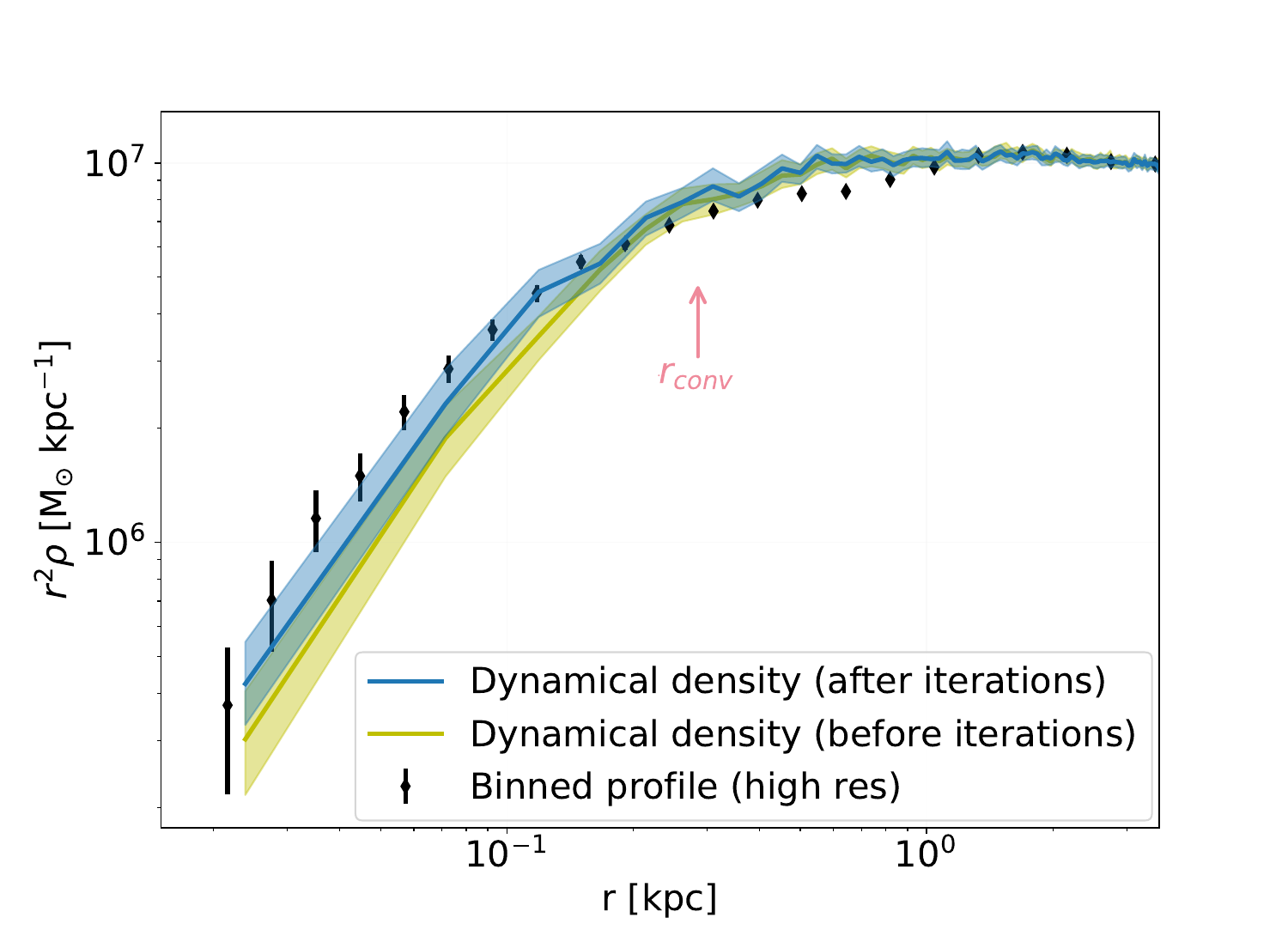}
    \caption{Dynamical density profile multiplied by $r^2$ before (yellow) and after (blue) the dynamical iteration process compared to the high resolution binned profile (black points), shown here for the example of Halo1459. The pink arrow marks the convergence radius of the low resolution simulation binned profile (which, for clarity, is not itself shown). The effect of the iterations is especially evident at small radii, where they act to make the central regions moderately denser, in better agreement with the high resolution profile.}
    \label{example_of_iteration}
\end{figure}

Figure \ref{example_of_iteration} shows the effect that the iteration process outlined has on the dynamical profile.
After the iterations, the profile's central gradient becomes moderately steeper. This can be understood by considering that the particles previously located at larger radii are now allowed to extend further inwards compared to their original positions in the snapshot, hence increasing the density in the inner regions. 
Note that the increase in central density may appear to violate mass conservation, since the total mass of the halo should be unaltered. However we verified that the mass enclosed converges to the same value at the virial radius; the volume of the sphere inside $r_{\mathrm{conv}}$ is just $3 \times 10^{-5} \%$ of the total volume inside the virial radius, and therefore a very small reduction in density across a large range of radii is able to provide the mass for an increased density cusp. 

Overall, we therefore conclude that the iterative component of the algorithm is important not just for self-consistency (as argued in Section $\S$\ref{potential_iteration_section}) but also to achieve the increased densities interior to the binned profile's convergence radius. Given that we kept actions fixed (to first order) during the iterations, one can envisage them as adiabatically transforming away some numerical effects of softening.

\subsubsection{Comparison at ultra-high resolution}\label{sec:ultra-high-res}

So far, we have applied our dynamical method to the \textit{low} resolution snapshots and compared our results against the binned profiles obtained from the high resolution versions of the simulations. In order to understand whether this improvement is independent of resolution, we now test the dynamical approach on the \textit{high} resolution simulations and compare the results to \textit{ultra-high} resolution snapshots. 

\begin{figure}
	\includegraphics[width=\columnwidth]{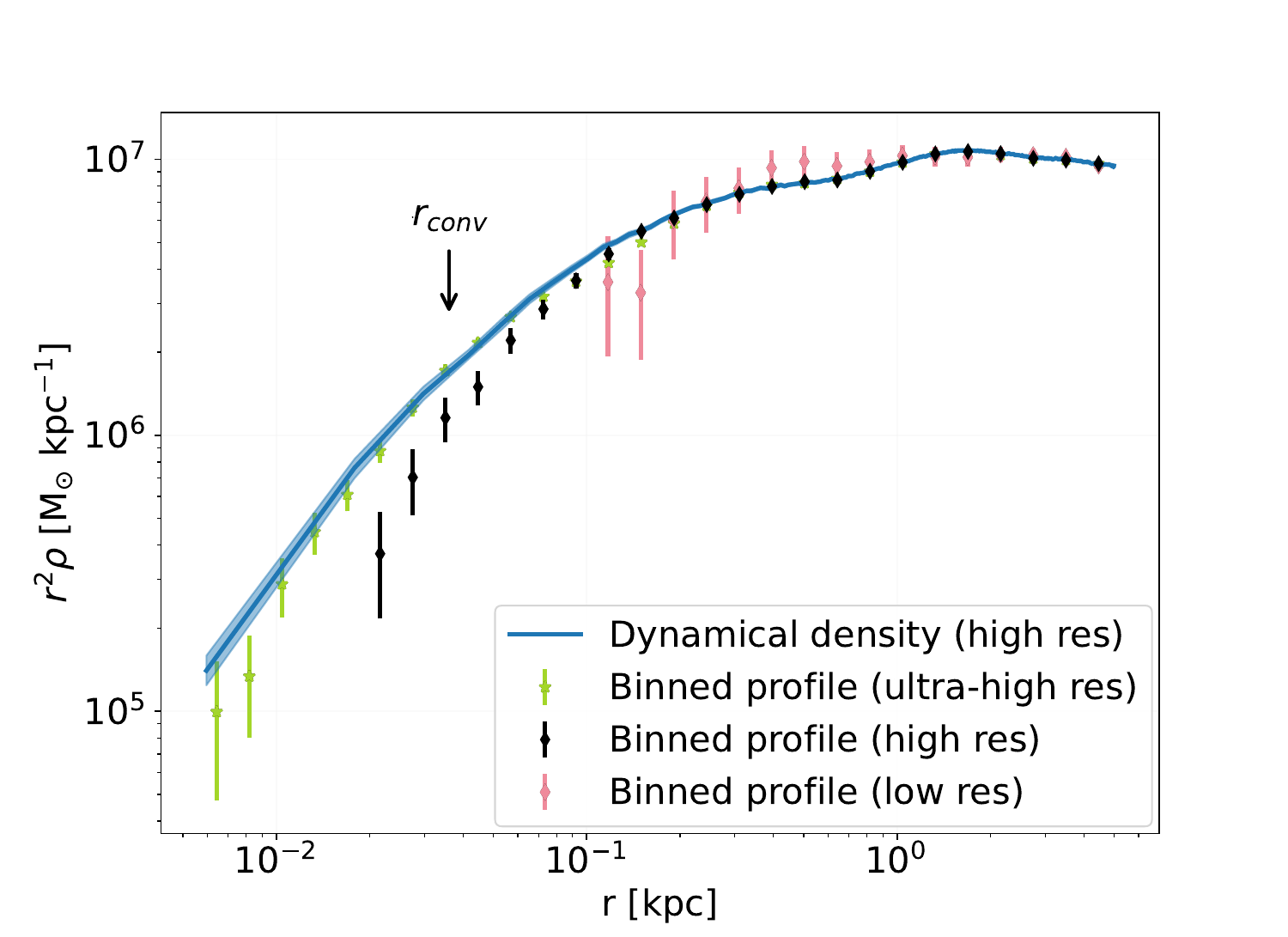}
    \caption{Dynamical density profile multiplied by $r^2$ (blue line) obtained from the \textit{high} resolution simulation of Halo1459, compared to the binned density profiles of the high (black points) and ultra-high (green points) resolution snapshots. The binned profile obtained from the low resolution snapshot is shown for reference (pink points). The black arrow indicates the approximate convergence radius of the high resolution binned profile ($3 \epsilon$). The dynamical density profile from the high resolution simulation predicts the ultra-high resolution simulation well, underscoring how the method can be applied at any resolution to extract additional information. }
    \label{higher_res_profile_extrapolation}
\end{figure}

Figure \ref{higher_res_profile_extrapolation} shows the dynamical density profile calculated from the high resolution simulation of Halo1459 compared to the binned distribution from an ultra-high resolution simulation with $\epsilon \simeq 6$ pc (half the softening length of the high resolution snapshots previously analysed). We take Halo1459 as an example, but similar results are observed for the other haloes.

All the conclusions drawn in the case of the low resolution dynamical profile are still valid when the code is applied to the high resolution snapshot: the dynamical density shows smaller uncertainties, a steeper cusp that extends further inwards and approximately follows the higher resolution binned profile, and a small density excess at $r{\sim}r_{\text{conv}}$ in the lower resolution profile. Overall, this confirms that the improvements obtained by adding dynamical information to the profiles continue even for increasingly precise simulations, making them resolution-independent.

\begin{figure}
	\includegraphics[width=\columnwidth]{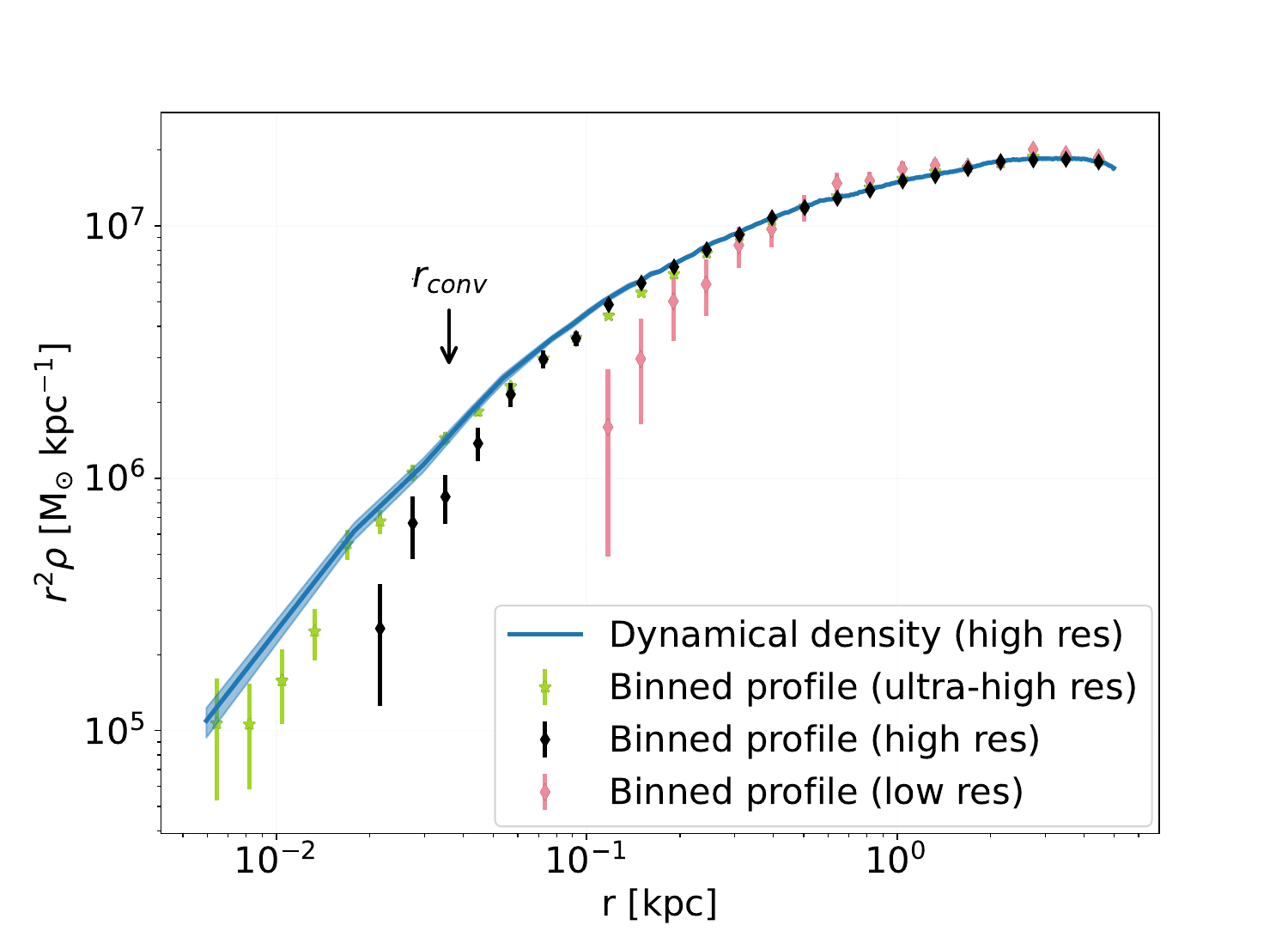}
    \caption{Same as Figure \ref{higher_res_profile_extrapolation} but for Halo600. The dynamical profile from the high resolution simulation of this halo shows a steep cusp consistent with the ultra-high resolution binned profile.
    The high resolution simulation, unlike the low resolution version, did not recently undergo a merger close to the halo's centre. This provides further evidence that the disagreement between the dynamical and binned profiles seen at small radii in the low resolution case is due to disequilibrium caused by the merger event.}
    \label{ultra-high_res_Halo600}
\end{figure}

In Figure \ref{ultra-high_res_Halo600} we show the dynamical profile obtained from the high resolution simulation of Halo600. 
When the dynamical code was previously applied to the low resolution simulation (top of Figure \ref{all_600s_halos}), we saw that the steepening in the cusp was insufficient to reach agreement with the high resolution binned profile. This is not the case when the dynamical profile is calculated from the high resolution snapshot: the cusp of the dynamical profile is entirely consistent with the ultra-high resolution binned profile.
This provides further evidence that the disagreement between the dynamical and binned profiles at small radii in the low resolution case is a result of the disequilibrium caused by the merger event, which did not occur in the high resolution version.

\subsection{Outer regions}\label{sec:outer-regions}

Having shown that the dynamical profile technique performs well in suppressing numerical noise at small radii (comparable to the convergence radius), we next consider its predictions at large radii (comparable to the virial radius $r_{\text{vir}}$). At such large radii, finite particle number is unlikely to be a limiting factor in drawing physical conclusions and therefore the motivation for studying the dynamical profile is different. Specifically, we are interested in understanding the degree to which halos may be considered equilibrium structures; departure from such equilibrium invalidates our assumptions and therefore should lead to an inaccurate profile. The virial radius roughly defines the point past which most particles are no longer gravitationally bound to the halo, such that infalling particles from the halo's environment begin to dominate.

We are able to study the dynamical profiles beyond  $r_{\text{vir}}$ for dwarf-scale haloes, since the zoom region extends several times further out. Beyond the virial radius we find, as expected, that the dynamical profiles are typically inaccurate; see Halo1445 and 1459 in Figure \ref{all_1400s_halos} for particularly clear cases. 
 
This provides one clear signature of out-of-equilibrium dynamics. However, another way to measure departures from equilibrium is via the binned average radial velocities of the particles ($\bar{v}_r$), which should be consistent with zero in equilibrium. Measured values of $\bar{v}_r$ are shown in the panels below the density profiles in Figures \ref{all_1400s_halos}, \ref{all_600s_halos}, and \ref{all_milkyway_halos}. As expected, these values deviate strongly from zero outside the virial radius, confirming our interpretation above. However, more surprisingly, the mean velocity values deviate from zero even {\it interior} to the virial radius, in regions where the binned and dynamical profiles fully agree (e.g. in Halo600, 605, 1459 over the radial range $1 < r < 40$ kpc). 
The root-mean square deviation of the radial velocities of all haloes (excluding Halo624) in the region $r < r_{\text{vir}}$ is of order $\sim 5\%$ of the virial velocity. 
These deviations are statistically important, and yet do not appear to have a significant effect on the overall density structure which is in good agreement with the binned estimates. This suggests that the dynamical profiles are robust to even significant violations of their equilibrium assumption.

\subsubsection{The role of substructures}

Although dynamical profiles remain robust despite the existence of smooth inflows detectable well interior to the virial radius, a more difficult challenge is posed by substructures. Most haloes have spikes in the {\it binned} density distribution at certain radii: for Halo600, 1445, and 1459 (Figure \ref{all_1400s_halos}, and top of Figure \ref{all_600s_halos}) these can be seen beyond the virial radius at $r \sim 90-100$ kpc, while for Halo624 (bottom of Figure \ref{all_600s_halos}) we see them much closer to the centre at $r \sim 10-20$ kpc. We refer to the locations of these features as $r_{\text{spike}}$. We verified that these local density spikes are indeed caused by substructures (see brown circles in the haloes density images in Figures \ref{all_1400s_halos}, \ref{all_600s_halos}, \ref{all_milkyway_halos}), which each contain between 3$\%$ and 9$\%$ of the mass of the main halo. All the other substructures present within the reflecting boundary have masses below 0.5$\%$ of the main halo's mass. 

The dynamical density profile does not reproduce spikes associated with substructure; by design, it smears them out along their orbit without taking into account the self-binding of the substructure. This leads to systematic differences between the binned and dynamical profiles, since the spike is smoothed out  while conserving the total mass. This effect is especially evident outside the virial radius in Halo1445 and 1459 (Figure \ref{all_1400s_halos}). In these cases, substructures (indicated by brown arrows at the appropriate radii on the density plots) coincide with significant disagreements between binned and dynamical halo profiles.

Halo624 contains a large substructure of mass ${\sim}1.4 \times 10^9 \text{M}\textsubscript{\(\odot\)}$ within its virial radius (at $r{\sim}$20-25 kpc). This is clearly visible in the density image at the bottom of Figure \ref{all_600s_halos}. The substructure will reach the centre of the main halo and merge in the next $\sim 500$ Myrs (based on its estimated infall velocity at $z=0$), and the disruption to the halo's equilibrium caused by the presence of substructure is also evident in the large deviations from zero in the average radial velocity panel. 
Despite this, the dynamical profile still faithfully represents the density distribution at radii between the centre of the halo and the location of the substructure. This shows that the effects of the dark matter spike are localised to the area around the substructure, and our method can represent the correct density distribution in other regions of the halo.

Halo605 provides an example with no large substructures present within the entire volume analysed. Despite fluctuations of the binned mean velocity, the dynamical profile agrees with the binned profile up to radii of $100\,\mathrm{kpc}$ which is around $2r_{\text{vir}}$. Taken with the discussion above, this counterexample strongly suggests that substructures, rather than smooth radial flows, are the dominant factor in determining whether binned and dynamical profiles differ significantly, and that the effect of substructures on the profile is always localised.

\subsubsection{Effect of the reflecting boundary}\label{bounding}

As described in Section $\S$\ref{create_dyn_densities}, the dynamical density profile requires an outer boundary condition. We have assumed a perfectly reflecting wall, which is equivalent to assuming that the particles flowing inwards across the boundary are exactly balanced by the flux outwards, in keeping with our broader assumption of dynamical equilibrium. However, there remains the freedom to move the reflecting wall to an arbitrary location. We carried out a number of experiments to determine the effect of this choice. If, for example, a boundary is placed inside the virial radius we found that the dynamical density profile is insensitive to the particular choice of location. However, in order to probe the outer parts of the halo the results above were all presented with the boundary outside the virial radius. In this case, there is more sensitivity to the particular choice of location.

\begin{figure}
	\includegraphics[width=\columnwidth]{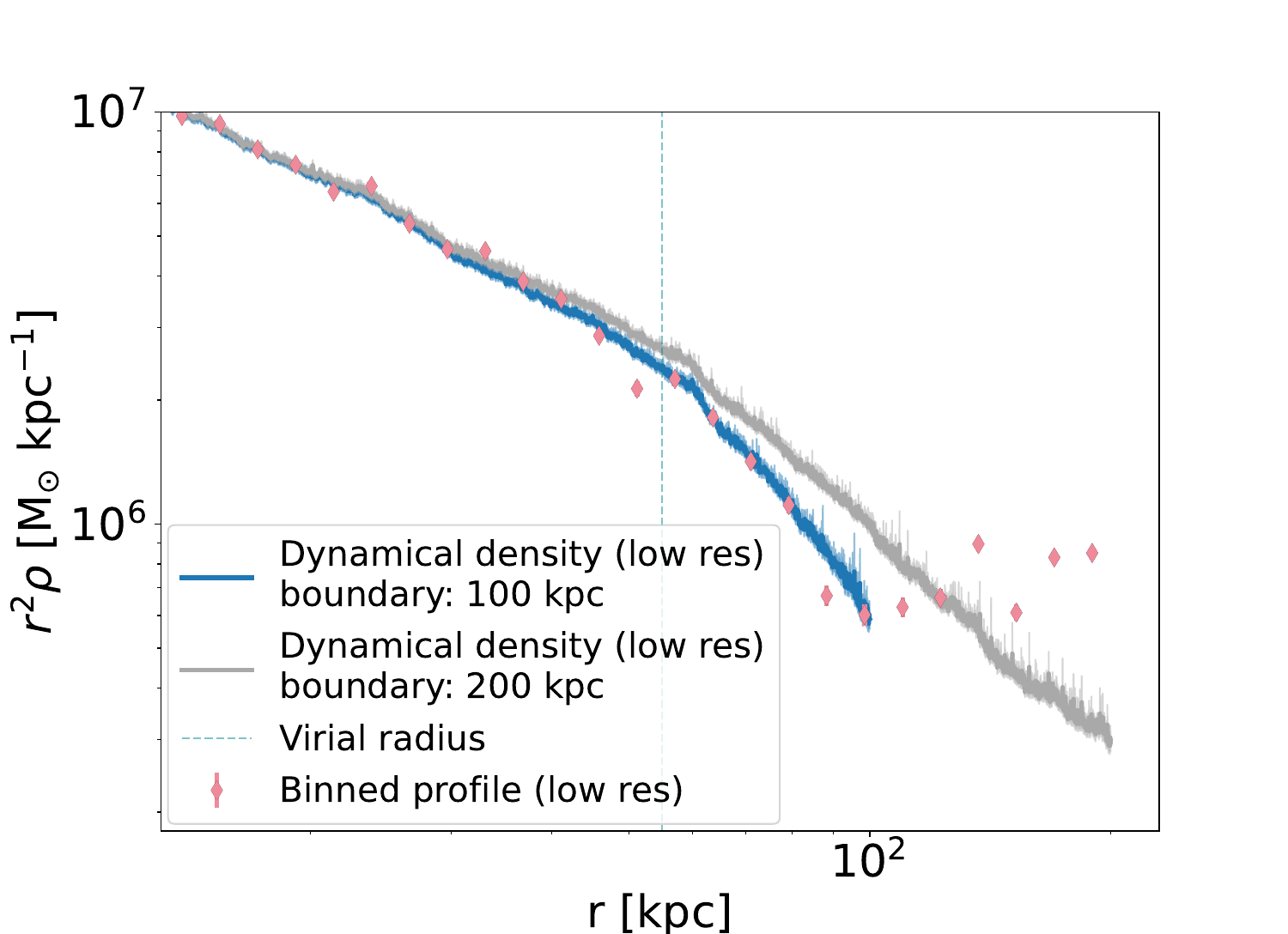}
    \caption{Zoom into the outer regions of the dynamical profile of Halo605 (middle of Figure \ref{all_600s_halos}) when the reflecting boundary is placed at 100 kpc (blue line) and then moved to 200 kpc (grey line) compared to the low resolution binned profile (pink points). The dynamical profile agrees well with the binned one when the boundary is placed anywhere up to 100 kpc, around twice the virial radius, but differs once contributions from particles out to 200 kpc are included in the calculations. These discrepancies propagate inwards to smaller radii, even below the virial radius (55 kpc, indicated by the vertical dashed line). This behaviour reflects our algorithm's extrapolation of how particles and substructures in the outskirts, while currently unbound, will ultimately fall into the halo at later times in the simulation, altering the density distribution.} 
    \label{change_profile_halo605}
\end{figure}

An example is shown in Figure \ref{change_profile_halo605} for Halo605. As usual, the binned profile is shown by pink points with error bars while dynamical profiles are represented as lines. Here, however, we show two alternative dynamical profiles: one with the reflecting boundary moved inwards to 100 kpc ($\simeq 2$ times the virial radius, as previously adopted, and illustrated here with a blue line) and one with the reflecting boundary moved outwards to 200 kpc ($\simeq 4$ times the virial radius, illustrated with a grey line). This shift causes the dynamical profile to deviate from the binned density in the range $r_{\text{vir}} < r < 2r_{\text{vir}}$, where there was previously agreement. 

The change is caused by particles that, at the time of the snapshot, are exterior to $2 r_{\text{vir}}$ but infalling, such that they spread to lower radii when the equilibrium assumption is imposed. The binned profile shows a `kink' at $\simeq 100\,\mathrm{kpc}$ which means that, in this particular case, there is a relatively large mass in such infalling particles. When the reflecting wall is located at $2r_{\text{vir}}$, these particles are safely isolated outside of the boundary, and therefore cannot affect the density profile.

In a sense, moving the reflecting wall to increasingly large radii provides a prediction of the future profile, since it extrapolates to a time when far-out particles have been able to fall into the inner regions. However, we did not study to what extent this can actually be used to make meaningful predictions and we caution that the actual process via which infalling particles relax into virial equilibrium is unlikely to be fully captured; in effect, our algorithm assumes conservation of their adiabatic invariants which is unlikely to be correct in detail.

For practical purposes, the most conservative choice of reflecting wall boundary is at the virial radius, but our  results show that it is entirely possible to obtain accurate profiles out to twice the virial radius. Beyond this, dynamical profiles with extended radial range may be of interest for understanding the accretion processes of halos and `splashback' features \citep{Diemer_2014,Adhikari_2014,More_2015,Shin_2023,LucieSmith22}, something we will investigate in the future.

\section{Conclusions and discussion} \label{discussion_section}

We presented a new method to estimate spherically-averaged densities in cosmological dark matter haloes. Instead of binning the particle in a snapshot by radius, which is the most obvious and prevalent approach, we use the velocity information in the snapshot to `smear' each particle along a trajectory, substantially reducing Poisson noise. 
Such a method has been proposed before \citep{Read_2005, Pontzen_2013}, but our work is the first systematic investigation of the approach. Additionally, we derive new corrections to take into account the integrable singularities at apocentre and pericentre, and introduce an iterative process to obtain a self-consistent potential-density pair. After iteration, we obtain central density estimates which (except in one case, Halo600, where a recent merger has occurred) follow the trend set by higher-resolution simulations. The agreement persists interior to the binned profile convergence radius, and all the way down to the simulation softening length. This highlights how our technique can squeeze extra information about the central regions of halos from existing simulations. 

In the outer regions, the dynamical profiles continue to agree with the binned profiles even out to several times the virial radius, provided that no substructures are present. If substructures are present, the assumption of equilibrium is locally broken and the profiles in the vicinity of the substructure are `smoothed' relative to the binned profiles. Nonetheless, the overall profiles remain accurate. Eventually, at approximately $r \sim 4 r_{\text{vir}} $, effects from the haloes' environments start to dominate, bringing the haloes too far out of equilibrium for the dynamical profiles to give meaningful density estimates. Including particles from these distant halo outskirts can produce changes to the dynamical profiles, sometimes even at radii below the virial radius. This is not a surprising result since the particles at large radii will eventually fall into the halo at future times in the simulation, and the dynamical approach is extrapolating the orbits of these particles accordingly. However, whether the resulting profile can be considered a `prediction' of the growth of the dark matter distribution at later times remains to be investigated.

These effects in the outer parts of the halo relate to the departure from perfect equilibrium (or phase-mixing), which is one of two key assumptions underlying the method. The second assumption is that the potential is spherically symmetric; this assumption is, in fact, broken by all our simulated halos, since they have triaxial equipotential surfaces. The fact that the dynamical profiles are accurate despite this broken assumption warrants further discussion.

\citet[][]{orkney2023edge} estimated the shapes of the five least massive dark matter haloes studied in this work by calculating the intermediate-to-major and minor-to-major axial ratios ($b/a$ and $c/a$) up to approximately 20 kpc in radius. The exact shape of each halo is not constant with radius: the $c/a$ ratio for all the haloes varies within the interval 0.4-0.8 (ratios of exactly 1 indicate perfect sphericity). The DMO haloes are generally the least spherical near their centre, becoming increasingly more spheroidal at radii beyond the cusp ($\gtrsim 1$ kpc).
Nevertheless, the dynamical density profiles are able to correctly represent the density distributions for the entire radial extent of the haloes.

The nature of the particles' orbits in an aspherical system is very different from the orbits that would be observed in a spherically-averaged version of the same potential. 
In the spherical case the angular momentum of individual particles is always constant; this is not the case in aspherical systems where only the total angular momentum of the entire system is conserved. This allows specific types of orbits (which would not be allowed in a spherical potential) to exist, such as box orbits which plunge through the centre of the halo.
Therefore, the fact that we are able to infer reliable results about the haloes' properties using only an artificial version of the dynamics which does not correspond to the real trajectories of the particles is not a straightforward outcome. 

However, such an outcome was previously predicted by relying on having a distribution function of particles in equilibrium \citep[][]{Pontzen_2015}. For every particle that is on an orbit losing angular momentum, there must be another particle on an orbit gaining angular momentum. To put it another way, the net flux of particles through the spherical action space must be everywhere zero, and so in a statistical sense, averaged across all particles, the spherical orbits remain a good approximation. For a more technical discussion, see \citet[][]{Pontzen_2015}. 
 The present work provides additional evidence that this mapping from a real triaxial system onto an effective spherical system is able to give accurate insights into dark matter halo structure. That said, the dynamical density method could be readily extended beyond the assumption of spherical symmetry, similarly to other mass modelling techniques \citep[][]{Schwarzschild_1979, Syer_1996}.

Overall, our dynamical method for the evaluation of dark matter density profiles is a powerful tool which can represent the correct mass distribution even when its fundamental assumptions are partially broken, making it largely applicable to a wide range of systems.

However, dark matter halos in the real universe have potentially been altered by baryonic effects, something which we have not investigated at all in the present paper. In forthcoming work, we will apply our dynamical density code to hydrodynamical simulations. Adding baryons to the simulations will likely alter the shape of the profile's inner regions, transforming the cusp into a flatter core. At a technical level, the gravitational potential can  no longer be made fully self-consistent with the dark matter density distribution, and the potential will need to be evaluated directly from the snapshot for the baryonic component. The iterative procedure that we have outlined will therefore need to be refined before we can use it in such cases.

\section*{Acknowledgements}

CM would like to thank the GMGalaxies team at UCL for useful discussions. 
CM is supported by the Science and Technology Facilities Council. 
AP is supported by the Royal Society.
JLS acknowledges the support of the Royal Society  (URF\textbackslash R1\textbackslash191555).
MPR is supported by the Beecroft Fellowship funded by Adrian Beecroft.
OA acknowledges support from the Knut and Alice Wallenberg Foundation and the Swedish Research Council (grant 2019-04659).
This study was supported by the European Research Council (ERC) under the European Union’s Horizon 2020 research and innovation programme (grant agreement No. 818085 GMGalaxies). This work was performed in part using the DiRAC Data Intensive service at Leicester, operated by the University of Leicester IT Services, which forms part of the STFC DiRAC HPC Facility (\url{www.dirac.ac.uk}). The equipment was funded by BEIS capital funding via STFC capital grants ST/K000373/1 and ST/R002363/1 and STFC DiRAC Operations grant ST/R001014/1. DiRAC is part of the National e-Infrastructure. This work was partially enabled by funding from the UCL Cosmoparticle Initiative.

\section*{Author Contributions}
The contributions from the authors are listed below using key-words based on the CRediT (Contribution Roles Taxonomy) system.

\textbf{CM:} investigation; methodology; software; formal analysis; visualisation; writing - original draft, review $\&$ editing.

\textbf{AP:} conceptualization; methodology; validation and interpretation; supervision; resources; writing - review $\&$ editing.

\textbf{JLS: } supervision; writing – review $\&$ editing.

\textbf{MPR: } data curation; writing – review $\&$ editing.

\textbf{JIR: } methodology; data curation; writing - review $\&$ editing.

\textbf{OA: } writing - review $\&$ editing.

\section*{Data Availability}

Data is available upon reasonable request. The code used to calculate the dynamical density profiles is publicly available on GitHub (repository: \href{https://github.com/claudiamuni/dynamical_density_profiles}{dynamical$\_$density$\_$profiles}).


\bibliographystyle{mnras}
\bibliography{Dynamical_density_paper}

\begin{thebibliography}{}
\makeatletter
\relax
\def\mn@urlcharsother{\let\do\@makeother \do\$\do\&\do\#\do\^\do\_\do\%\do\~}
\def\mn@doi{\begingroup\mn@urlcharsother \@ifnextchar [ {\mn@doi@} {\mn@doi@[]}}
\def\mn@doi@[#1]#2{\def\@tempa{#1}\ifx\@tempa\@empty \href {http://dx.doi.org/#2} {doi:#2}\else \href {http://dx.doi.org/#2} {#1}\fi \endgroup}
\def\mn@eprint#1#2{\mn@eprint@#1:#2::\@nil}
\def\mn@eprint@arXiv#1{\href {http://arxiv.org/abs/#1} {{\tt arXiv:#1}}}
\def\mn@eprint@dblp#1{\href {http://dblp.uni-trier.de/rec/bibtex/#1.xml} {dblp:#1}}
\def\mn@eprint@#1:#2:#3:#4\@nil{\def\@tempa {#1}\def\@tempb {#2}\def\@tempc {#3}\ifx \@tempc \@empty \let \@tempc \@tempb \let \@tempb \@tempa \fi \ifx \@tempb \@empty \def\@tempb {arXiv}\fi \@ifundefined {mn@eprint@\@tempb}{\@tempb:\@tempc}{\expandafter \expandafter \csname mn@eprint@\@tempb\endcsname \expandafter{\@tempc}}}

\bibitem[\protect\citeauthoryear{Adhikari, Dalal  \& Chamberlain}{Adhikari et~al.}{2014}]{Adhikari_2014}
Adhikari S.,  Dalal N.,   Chamberlain R.~T.,  2014, \mn@doi [Journal of Cosmology and Astroparticle Physics] {10.1088/1475-7516/2014/11/019}, 2014, 019

\bibitem[\protect\citeauthoryear{Agertz et~al.,}{Agertz et~al.}{2019}]{Agertz_2019}
Agertz O.,  et~al., 2019, \mn@doi [MNRAS] {10.1093/mnras/stz3053}, 491, 1656

\bibitem[\protect\citeauthoryear{Allgood, Flores, Primack, Kravtsov, Wechsler, Faltenbacher  \& Bullock}{Allgood et~al.}{2006}]{Allgood_2006}
Allgood B.,  Flores R.~A.,  Primack J.~R.,  Kravtsov A.~V.,  Wechsler R.~H.,  Faltenbacher A.,   Bullock J.~S.,  2006, \mn@doi [MNRAS] {10.1111/j.1365-2966.2006.10094.x}, 367, 1781

\bibitem[\protect\citeauthoryear{Angulo \& Hahn}{Angulo \& Hahn}{2022}]{Angulo_2022}
Angulo R.~E.,  Hahn O.,  2022, \mn@doi [Living Reviews in Computational Astrophysics] {10.1007/s41115-021-00013-z}, 8

\bibitem[\protect\citeauthoryear{Battaglia, Helmi, Tolstoy, Irwin, Hill  \& Jablonka}{Battaglia et~al.}{2008}]{Battaglia_2008}
Battaglia G.,  Helmi A.,  Tolstoy E.,  Irwin M.,  Hill V.,   Jablonka P.,  2008, \mn@doi [ApJ] {10.1086/590179}, 681, L13

\bibitem[\protect\citeauthoryear{Burkert}{Burkert}{1995}]{Burkert_1995}
Burkert A.,  1995, \mn@doi [ApJ] {10.1086/309560}, 447

\bibitem[\protect\citeauthoryear{Callingham, Cautun, Deason, Frenk, Grand, Marinacci  \& Pakmor}{Callingham et~al.}{2020}]{Callingham_2020}
Callingham T.~M.,  Cautun M.,  Deason A.~J.,  Frenk C.~S.,  Grand R. J.~J.,  Marinacci F.,   Pakmor R.,  2020, \mn@doi [MNRAS] {10.1093/mnras/staa1089}, 495, 12–28

\bibitem[\protect\citeauthoryear{Chemin, de Blok  \& Mamon}{Chemin et~al.}{2011}]{Chemin_2011}
Chemin L.,  de Blok W. J.~G.,   Mamon G.~A.,  2011, \mn@doi [AJ] {10.1088/0004-6256/142/4/109}, 142, 109

\bibitem[\protect\citeauthoryear{{De Leo}, Read, Noel, Erkal, Massana  \& Carrera}{{De Leo} et~al.}{2023}]{DeLeo_2023}
{De Leo} M.,  Read J.~I.,  Noel N. E.~D.,  Erkal D.,  Massana P.,   Carrera R.,  2023, Surviving the Waves: evidence for a Dark Matter cusp in the tidally disrupting Small Magellanic Cloud (\mn@eprint {arXiv} {2303.08838})

\bibitem[\protect\citeauthoryear{Dehnen \& Read}{Dehnen \& Read}{2011}]{Dehnen_2011}
Dehnen W.,  Read J.~I.,  2011, \mn@doi [The European Physical Journal Plus] {10.1140/epjp/i2011-11055-3}, 126

\bibitem[\protect\citeauthoryear{Diemand, Moore  \& Stadel}{Diemand et~al.}{2004}]{Diemand_2004}
Diemand J.,  Moore B.,   Stadel J.,  2004, \mn@doi [MNRAS] {10.1111/j.1365-2966.2004.08094.x}, 353, 624

\bibitem[\protect\citeauthoryear{Diemer}{Diemer}{2022a}]{Diemer_2022}
Diemer B.,  2022a, \mn@doi [MNRAS] {10.1093/mnras/stac878}, 513, 573

\bibitem[\protect\citeauthoryear{Diemer}{Diemer}{2022b}]{Diemer_2022b}
Diemer B.,  2022b, \mn@doi [MNRAS] {10.1093/mnras/stac3778}, 519, 3292

\bibitem[\protect\citeauthoryear{Diemer \& Kravtsov}{Diemer \& Kravtsov}{2014}]{Diemer_2014}
Diemer B.,  Kravtsov A.~V.,  2014, \mn@doi [ApJ] {10.1088/0004-637x/789/1/1}, 789, 1

\bibitem[\protect\citeauthoryear{{Dubinski} \& {Carlberg}}{{Dubinski} \& {Carlberg}}{1991}]{Dubinski_1991}
{Dubinski} J.,  {Carlberg} R.~G.,  1991, \mn@doi [\apj] {10.1086/170451}, \href {https://ui.adsabs.harvard.edu/abs/1991ApJ...378..496D} {378, 496}

\bibitem[\protect\citeauthoryear{Dutton \& Macci{\`{o} }}{Dutton \& Macci{\`{o} }}{2014}]{Dutton_2014}
Dutton A.~A.,  Macci{\`{o} } A.~V.,  2014, \mn@doi [MNRAS] {10.1093/mnras/stu742}, 441, 3359

\bibitem[\protect\citeauthoryear{{Einasto}}{{Einasto}}{1965}]{Einasto_1965}
{Einasto} J.,  1965, Trudy Astrofizicheskogo Instituta Alma-Ata, \href {https://ui.adsabs.harvard.edu/abs/1965TrAlm...5...87E} {5, 87}

\bibitem[\protect\citeauthoryear{El-Zant, Shlosman  \& Hoffman}{El-Zant et~al.}{2001}]{El-Zant_2001}
El-Zant A.,  Shlosman I.,   Hoffman Y.,  2001, \mn@doi [ApJ] {10.1086/322516}, 560, 636

\bibitem[\protect\citeauthoryear{Fielder, Mao, Zentner, Newman, Wu  \& Wechsler}{Fielder et~al.}{2020}]{Fielder_2020}
Fielder C.~E.,  Mao Y.-Y.,  Zentner A.~R.,  Newman J.~A.,  Wu H.-Y.,   Wechsler R.~H.,  2020, \mn@doi [MNRAS] {10.1093/mnras/staa2851}, 499, 2426

\bibitem[\protect\citeauthoryear{Flores \& Primack}{Flores \& Primack}{1994}]{Flores_1994}
Flores R.~A.,  Primack J.~R.,  1994, \mn@doi [ApJ] {10.1086/187350}, 427, L1

\bibitem[\protect\citeauthoryear{Frenk \& White}{Frenk \& White}{2012}]{Frenk_2012}
Frenk C.,  White S.,  2012, \mn@doi [Annalen der Physik] {10.1002/andp.201200212}, 524, 507

\bibitem[\protect\citeauthoryear{{Frenk}, {White}, {Davis}  \& {Efstathiou}}{{Frenk} et~al.}{1988}]{Frenk_1988}
{Frenk} C.~S.,  {White} S. D.~M.,  {Davis} M.,   {Efstathiou} G.,  1988, \mn@doi [\apj] {10.1086/166213}, \href {https://ui.adsabs.harvard.edu/abs/1988ApJ...327..507F} {327, 507}

\bibitem[\protect\citeauthoryear{Fukushige \& Makino}{Fukushige \& Makino}{2001}]{Fukushige_2001}
Fukushige T.,  Makino J.,  2001, \mn@doi [ApJ] {10.1086/321666}, 557, 533

\bibitem[\protect\citeauthoryear{Gao, Navarro, Frenk, Jenkins, Springel  \& White}{Gao et~al.}{2012}]{Gao_2012}
Gao L.,  Navarro J.~F.,  Frenk C.~S.,  Jenkins A.,  Springel V.,   White S. D.~M.,  2012, \mn@doi [MNRAS] {10.1111/j.1365-2966.2012.21564.x}, 425, 2169

\bibitem[\protect\citeauthoryear{Genina et~al.,}{Genina et~al.}{2017}]{Genina_2017}
Genina A.,  et~al., 2017, \mn@doi [MNRAS] {10.1093/mnras/stx2855}, 474, 1398

\bibitem[\protect\citeauthoryear{Ghigna, Moore, Governato, Lake, Quinn  \& Stadel}{Ghigna et~al.}{2000}]{Ghigna_2000}
Ghigna S.,  Moore B.,  Governato F.,  Lake G.,  Quinn T.,   Stadel J.,  2000, \mn@doi [ApJ] {10.1086/317221}, 544, 616

\bibitem[\protect\citeauthoryear{Hague \& Wilkinson}{Hague \& Wilkinson}{2013}]{Hague_2013}
Hague P.~R.,  Wilkinson M.~I.,  2013, \mn@doi [MNRAS] {10.1093/mnras/stt899}, 433, 2314

\bibitem[\protect\citeauthoryear{Hayashi, Chiba  \& Ishiyama}{Hayashi et~al.}{2020}]{Hayashi_2020}
Hayashi K.,  Chiba M.,   Ishiyama T.,  2020, \mn@doi [ApJ] {10.3847/1538-4357/abbe0a}, 904, 45

\bibitem[\protect\citeauthoryear{{Hernquist}}{{Hernquist}}{1990}]{Hernquist_1990}
{Hernquist} L.,  1990, \mn@doi [\apj] {10.1086/168845}, \href {https://ui.adsabs.harvard.edu/abs/1990ApJ...356..359H} {356, 359}

\bibitem[\protect\citeauthoryear{{Jing} \& {Suto}}{{Jing} \& {Suto}}{2002}]{Jing_2002}
{Jing} Y.~P.,  {Suto} Y.,  2002, \mn@doi [\apj] {10.1086/341065}, \href {https://ui.adsabs.harvard.edu/abs/2002ApJ...574..538J} {574, 538}

\bibitem[\protect\citeauthoryear{{Lucie-Smith}, {Peiris}, {Pontzen}, {Nord}, {Thiyagalingam}  \& {Piras}}{{Lucie-Smith} et~al.}{2022}]{LucieSmith22}
{Lucie-Smith} L.,  {Peiris} H.~V.,  {Pontzen} A.,  {Nord} B.,  {Thiyagalingam} J.,   {Piras} D.,  2022, \mn@doi [\prd] {10.1103/PhysRevD.105.103533}, \href {https://ui.adsabs.harvard.edu/abs/2022PhRvD.105j3533L} {105, 103533}

\bibitem[\protect\citeauthoryear{Ludlow, Schaye  \& Bower}{Ludlow et~al.}{2019}]{Ludlow_2019}
Ludlow A.~D.,  Schaye J.,   Bower R.,  2019, \mn@doi [MNRAS] {10.1093/mnras/stz1821}, 488, 3663

\bibitem[\protect\citeauthoryear{Marchesini, D'Onghia, Chincarini, Firmani, Conconi, Molinari  \& Zacchei}{Marchesini et~al.}{2002}]{Marchesini_2002}
Marchesini D.,  D'Onghia E.,  Chincarini G.,  Firmani C.,  Conconi P.,  Molinari E.,   Zacchei A.,  2002, \mn@doi [ApJ] {10.1086/341475}, 575, 801

\bibitem[\protect\citeauthoryear{More, Diemer  \& Kravtsov}{More et~al.}{2015}]{More_2015}
More S.,  Diemer B.,   Kravtsov A.~V.,  2015, \mn@doi [ApJ] {10.1088/0004-637x/810/1/36}, 810, 36

\bibitem[\protect\citeauthoryear{Navarro, Eke  \& Frenk}{Navarro et~al.}{1996a}]{Navarro_1996a}
Navarro J.~F.,  Eke V.~R.,   Frenk C.~S.,  1996a, \mn@doi [MNRAS] {10.1093/mnras/283.3.l72}, 283, L72

\bibitem[\protect\citeauthoryear{Navarro, Frenk  \& White}{Navarro et~al.}{1996b}]{Navarro_1996}
Navarro J.~F.,  Frenk C.~S.,   White S. D.~M.,  1996b, \mn@doi [ApJ] {10.1086/177173}, 462, 563

\bibitem[\protect\citeauthoryear{Navarro, Frenk  \& White}{Navarro et~al.}{1997}]{Navarro_1997}
Navarro J.~F.,  Frenk C.~S.,   White S. D.~M.,  1997, \mn@doi [ApJ] {10.1086/304888}, 490, 493

\bibitem[\protect\citeauthoryear{Navarro et~al.,}{Navarro et~al.}{2004}]{Navarro_2004}
Navarro J.~F.,  et~al., 2004, \mn@doi [MNRAS] {10.1111/j.1365-2966.2004.07586.x}, 349, 1039

\bibitem[\protect\citeauthoryear{Nipoti \& Binney}{Nipoti \& Binney}{2014}]{Nipoti_2014}
Nipoti C.,  Binney J.,  2014, \mn@doi [MNRAS] {10.1093/mnras/stu2217}, 446, 1820

\bibitem[\protect\citeauthoryear{Oh et~al.,}{Oh et~al.}{2015}]{Oh_2015}
Oh S.-H.,  et~al., 2015, \mn@doi [AJ] {10.1088/0004-6256/149/6/180}, 149, 180

\bibitem[\protect\citeauthoryear{Oldham \& Auger}{Oldham \& Auger}{2016}]{Oldham_2016}
Oldham L.~J.,  Auger M.~W.,  2016, \mn@doi [MNRAS] {10.1093/mnras/stv2982}, 457, 421

\bibitem[\protect\citeauthoryear{Oman, Marasco, Navarro, Frenk, Schaye  \& Ben{\'{\i} }tez-Llambay}{Oman et~al.}{2018}]{Oman_2018}
Oman K.~A.,  Marasco A.,  Navarro J.~F.,  Frenk C.~S.,  Schaye J.,   Ben{\'{\i} }tez-Llambay A.,  2018, \mn@doi [MNRAS] {10.1093/mnras/sty2687}, 482, 821

\bibitem[\protect\citeauthoryear{Orkney et~al.,}{Orkney et~al.}{2021}]{Orkney_2021}
Orkney M. D.~A.,  et~al., 2021, \mn@doi [MNRAS] {10.1093/mnras/stab1066}, 504, 3509

\bibitem[\protect\citeauthoryear{Orkney et~al.,}{Orkney et~al.}{2022}]{Orkney_2022}
Orkney M. D.~A.,  et~al., 2022, \mn@doi [MNRAS] {10.1093/mnras/stac1755}, 515, 185

\bibitem[\protect\citeauthoryear{Orkney, Taylor, Read, Rey, Pontzen, Agertz, Kim  \& Delorme}{Orkney et~al.}{2023}]{Orkney_2023}
Orkney M. D.~A.,  Taylor E.,  Read J.~I.,  Rey M.~P.,  Pontzen A.,  Agertz O.,  Kim S.~Y.,   Delorme M.,  2023, \mn@doi [Monthly Notices of the Royal Astronomical Society] {10.1093/mnras/stad2516}, 525, 3516–3532

\bibitem[\protect\citeauthoryear{Pineda, Hayward, Springel  \& Mendes~de Oliveira}{Pineda et~al.}{2016}]{Pineda_2016}
Pineda J. C.~B.,  Hayward C.~C.,  Springel V.,   Mendes~de Oliveira C.,  2016, \mn@doi [MNRAS] {10.1093/mnras/stw3004}, 466, 63

\bibitem[\protect\citeauthoryear{{Planck Collaboration} et~al.,}{{Planck Collaboration} et~al.}{2014}]{Plank_2014}
{Planck Collaboration} et~al., 2014, \mn@doi [A\&A] {10.1051/0004-6361/201321591}, 571, A16

\bibitem[\protect\citeauthoryear{{Planck Collaboration} et~al.,}{{Planck Collaboration} et~al.}{2016}]{Plank_2016}
{Planck Collaboration} et~al., 2016, \mn@doi [A\&A] {10.1051/0004-6361/201525830}, 594, A13

\bibitem[\protect\citeauthoryear{Pontzen \& Governato}{Pontzen \& Governato}{2012}]{Pontzen_2012}
Pontzen A.,  Governato F.,  2012, \mn@doi [MNRAS] {10.1111/j.1365-2966.2012.20571.x}, 421, 3464

\bibitem[\protect\citeauthoryear{Pontzen \& Governato}{Pontzen \& Governato}{2013}]{Pontzen_2013}
Pontzen A.,  Governato F.,  2013, \mn@doi [MNRAS] {10.1093/mnras/sts529}, 430, 121

\bibitem[\protect\citeauthoryear{{Pontzen}, {Ro{\v{s}}kar}, {Stinson}  \& {Woods}}{{Pontzen} et~al.}{2013}]{pynbody}
{Pontzen} A.,  {Ro{\v{s}}kar} R.,  {Stinson} G.,   {Woods} R.,  2013, {pynbody: N-Body/SPH analysis for python}, Astrophysics Source Code Library, record ascl:1305.002 (\mn@eprint {ascl} {1305.002})

\bibitem[\protect\citeauthoryear{Pontzen, Read, Teyssier, Governato, Gualandris, Roth  \& Devriendt}{Pontzen et~al.}{2015}]{Pontzen_2015}
Pontzen A.,  Read J.~I.,  Teyssier R.,  Governato F.,  Gualandris A.,  Roth N.,   Devriendt J.,  2015, \mn@doi [MNRAS] {10.1093/mnras/stv1032}, 451, 1366

\bibitem[\protect\citeauthoryear{Popolo \& Pace}{Popolo \& Pace}{2016}]{Del_Popolo_2016}
Popolo A.~D.,  Pace F.,  2016, \mn@doi [Ap&SS] {10.1007/s10509-016-2742-z}, 361

\bibitem[\protect\citeauthoryear{Power, Navarro, Jenkins, Frenk, White, Springel, Stadel  \& Quinn}{Power et~al.}{2003}]{Power_2003}
Power C.,  Navarro J.~F.,  Jenkins A.,  Frenk C.~S.,  White S. D.~M.,  Springel V.,  Stadel J.,   Quinn T.,  2003, \mn@doi [MNRAS] {10.1046/j.1365-8711.2003.05925.x}, 338, 14

\bibitem[\protect\citeauthoryear{Read \& Gilmore}{Read \& Gilmore}{2005}]{Read_2005}
Read J.~I.,  Gilmore G.,  2005, \mn@doi [MNRAS] {10.1111/j.1365-2966.2004.08424.x}, 356, 107

\bibitem[\protect\citeauthoryear{Read, Agertz  \& Collins}{Read et~al.}{2016}]{Read_2016}
Read J.~I.,  Agertz O.,   Collins M. L.~M.,  2016, \mn@doi [MNRAS] {10.1093/mnras/stw713}, 459, 2573

\bibitem[\protect\citeauthoryear{Read, Iorio, Agertz  \& Fraternali}{Read et~al.}{2017}]{Read_2017}
Read J.~I.,  Iorio G.,  Agertz O.,   Fraternali F.,  2017, \mn@doi [MNRAS] {10.1093/mnras/stx147}, 467, 2019

\bibitem[\protect\citeauthoryear{Read, Walker  \& Steger}{Read et~al.}{2019}]{Read_2019}
Read J.~I.,  Walker M.~G.,   Steger P.,  2019, \mn@doi [MNRAS] {10.1093/mnras/sty3404}, 484, 1401

\bibitem[\protect\citeauthoryear{Rey \& Starkenburg}{Rey \& Starkenburg}{2021}]{Rey_2021_vintergatan}
Rey M.~P.,  Starkenburg T.~K.,  2021, \mn@doi [MNRAS] {10.1093/mnras/stab3709}, 510, 4208

\bibitem[\protect\citeauthoryear{Rey, Pontzen, Agertz, Orkney, Read, Saintonge  \& Pedersen}{Rey et~al.}{2019}]{Rey_2019}
Rey M.~P.,  Pontzen A.,  Agertz O.,  Orkney M. D.~A.,  Read J.~I.,  Saintonge A.,   Pedersen C.,  2019, \mn@doi [ApJ Letters] {10.3847/2041-8213/ab53dd}, 886, L3

\bibitem[\protect\citeauthoryear{Rey, Pontzen, Agertz, Orkney, Read  \& Rosdahl}{Rey et~al.}{2020}]{Rey_2020}
Rey M.~P.,  Pontzen A.,  Agertz O.,  Orkney M. D.~A.,  Read J.~I.,   Rosdahl J.,  2020, \mn@doi [MNRAS] {10.1093/mnras/staa1640}, 497, 1508

\bibitem[\protect\citeauthoryear{Salucci, Lapi, Tonini, Gentile, Yegorova  \& Klein}{Salucci et~al.}{2007}]{Salucci_2007}
Salucci P.,  Lapi A.,  Tonini C.,  Gentile G.,  Yegorova I.,   Klein U.,  2007, \mn@doi [MNRAS] {10.1111/j.1365-2966.2007.11696.x}, 378, 41

\bibitem[\protect\citeauthoryear{{Schwarzschild}}{{Schwarzschild}}{1979}]{Schwarzschild_1979}
{Schwarzschild} M.,  1979, \mn@doi [\apj] {10.1086/157282}, \href {https://ui.adsabs.harvard.edu/abs/1979ApJ...232..236S} {232, 236}

\bibitem[\protect\citeauthoryear{Shin \& Diemer}{Shin \& Diemer}{2023}]{Shin_2023}
Shin T.,  Diemer B.,  2023, \mn@doi [MNRAS] {10.1093/mnras/stad860}, 521, 5570

\bibitem[\protect\citeauthoryear{{Shin} et~al.,}{{Shin} et~al.}{2019}]{Shin19Splashback}
{Shin} T.,  et~al., 2019, \mn@doi [\mnras] {10.1093/mnras/stz1434}, \href {https://ui.adsabs.harvard.edu/abs/2019MNRAS.487.2900S} {487, 2900}

\bibitem[\protect\citeauthoryear{{Syer} \& {Tremaine}}{{Syer} \& {Tremaine}}{1996}]{Syer_1996}
{Syer} D.,  {Tremaine} S.,  1996, \mn@doi [\mnras] {10.1093/mnras/282.1.223}, \href {https://ui.adsabs.harvard.edu/abs/1996MNRAS.282..223S} {282, 223}

\bibitem[\protect\citeauthoryear{{Teyssier}}{{Teyssier}}{2002}]{Teyssier_2002}
{Teyssier} R.,  2002, \mn@doi [\aap] {10.1051/0004-6361:20011817}, \href {https://ui.adsabs.harvard.edu/abs/2002A&A...385..337T} {385, 337}

\bibitem[\protect\citeauthoryear{Vasiliev}{Vasiliev}{2014}]{Vasiliev_2014}
Vasiliev E.,  2014, \mn@doi [MNRAS] {10.1093/mnras/stu2360}, 446, 3150–3161

\bibitem[\protect\citeauthoryear{Vogelsberger, Marinacci, Torrey  \& Puchwein}{Vogelsberger et~al.}{2020}]{Vogelsberger_2020}
Vogelsberger M.,  Marinacci F.,  Torrey P.,   Puchwein E.,  2020, \mn@doi [Nature Reviews Physics] {10.1038/s42254-019-0127-2}, 2, 42

\bibitem[\protect\citeauthoryear{Walker \& Pe{\~n}arrubia}{Walker \& Pe{\~n}arrubia}{2011}]{Walker_2011}
Walker M.~G.,  Pe{\~n}arrubia J.,  2011, \mn@doi [ApJ] {10.1088/0004-637X/742/1/20}, 742, 20

\bibitem[\protect\citeauthoryear{{Wang}, {Bose}, {Frenk}, {Gao}, {Jenkins}, {Springel}  \& {White}}{{Wang} et~al.}{2020}]{Wang_2020_N}
{Wang} J.,  {Bose} S.,  {Frenk} C.~S.,  {Gao} L.,  {Jenkins} A.,  {Springel} V.,   {White} S.~D.~M.,  2020, \mn@doi [\nat] {10.1038/s41586-020-2642-9}, \href {https://ui.adsabs.harvard.edu/abs/2020Natur.585...39W} {585, 39}

\bibitem[\protect\citeauthoryear{Zhao}{Zhao}{1996}]{Zhao_1996}
Zhao H.,  1996, \mn@doi [MNRAS] {10.1093/mnras/278.2.488}, 278, 488

\bibitem[\protect\citeauthoryear{{Zoutendijk}, {Brinchmann}, {Bouch{\'e}}, {den Brok}, {Krajnovi{\'c}}, {Kuijken}, {Maseda}  \& {Schaye}}{{Zoutendijk} et~al.}{2021}]{Zoutendijk_2021}
{Zoutendijk} S.~L.,  {Brinchmann} J.,  {Bouch{\'e}} N.~F.,  {den Brok} M.,  {Krajnovi{\'c}} D.,  {Kuijken} K.,  {Maseda} M.~V.,   {Schaye} J.,  2021, \mn@doi [\aap] {10.1051/0004-6361/202040239}, \href {https://ui.adsabs.harvard.edu/abs/2021A&A...651A..80Z} {651, A80}

\bibitem[\protect\citeauthoryear{de Blok, McGaugh  \& Rubin}{de~Blok et~al.}{2001}]{deBlok_2001}
de Blok W. J.~G.,  McGaugh S.~S.,   Rubin V.~C.,  2001, \mn@doi [AJ] {10.1086/323450}, 122, 2396

\makeatother
\end{thebibliography}






\bsp	
\label{lastpage}
\end{document}